\documentstyle[epsf,rotate]{mn}
\newif\ifAMStwofonts
\AMStwofontstrue

\def\O{$\Omega$}
\input wasyfont
 
\def\ar{{$\rightarrow$}}

\def\D{{$\Delta$ }}           

\ifoldfss
  \ifCUPmtlplainloaded \else
    \NewTextAlphabet{textbfit} {cmbxti10} {}
    \NewTextAlphabet{textbfss} {cmssbx10} {}
    \NewMathAlphabet{mathbfit} {cmbxti10} {} 
    \NewMathAlphabet{mathbfss} {cmssbx10} {} 
  \fi
  \ifAMStwofonts
    \ifCUPmtlplainloaded \else
      \NewSymbolFont{upmath} {eurm10}
      \NewSymbolFont{AMSa} {msam10}
      \NewMathSymbol{\upi}     {0}{upmath}{19}
      \NewMathSymbol{\umu}     {0}{upmath}{16}
      \NewMathSymbol{\upartial}{0}{upmath}{40}
      \NewMathSymbol{\leqslant}{3}{AMSa}{36}
      \NewMathSymbol{\geqslant}{3}{AMSa}{3E}

    \fi
  \fi
\fi 

\ifnfssone
  \newmathalphabet{\mathit}
  \addtoversion{normal}{\mathit}{cmr}{m}{it}
  \addtoversion{bold}{\mathit}{cmr}{bx}{it}
  \newmathalphabet{\mathbfit} 
  \addtoversion{normal}{\mathbfit}{cmr}{bx}{it}
  \addtoversion{bold}{\mathbfit}{cmr}{bx}{it}
  \newmathalphabet{\mathbfss} 
  \addtoversion{normal}{\mathbfss}{cmss}{bx}{n}
  \addtoversion{bold}{\mathbfss}{cmss}{bx}{n}
  \ifAMStwofonts
    \ifCUPmtlplainloaded \else
      \UseAMStwoboldmath
      \makeatletter
      \new@mathgroup\upmath@group
      \define@mathgroup\mv@normal\upmath@group{eur}{m}{n}
      \define@mathgroup\mv@bold\upmath@group{eur}{b}{n}
      \edef\UPM{\hexnumber\upmath@group}
      \new@mathgroup\amsa@group
      \define@mathgroup\mv@normal\amsa@group{msa}{m}{n}
      \define@mathgroup\mv@bold\amsa@group{msa}{m}{n}
      \edef\AMSa{\hexnumber\amsa@group}
      \makeatother
      \mathchardef\upi="0\UPM19
      \mathchardef\umu="0\UPM16
      \mathchardef\upartial="0\UPM40
      \mathchardef\leqslant="3\AMSa36
      \mathchardef\geqslant="3\AMSa3E
    \fi
  \fi
\fi 

\ifnfsstwo
  \DeclareMathAlphabet{\mathbfit}{OT1}{cmr}{bx}{it}
  \SetMathAlphabet\mathbfit{bold}{OT1}{cmr}{bx}{it}
  \DeclareMathAlphabet{\mathbfss}{OT1}{cmss}{bx}{n}
  \SetMathAlphabet\mathbfss{bold}{OT1}{cmss}{bx}{n}
  \ifAMStwofonts
    \ifCUPmtlplainloaded \else
      \DeclareSymbolFont{UPM}{U}{eur}{m}{n}
      \SetSymbolFont{UPM}{bold}{U}{eur}{b}{n}
      \DeclareSymbolFont{AMSa}{U}{msa}{m}{n}
      \DeclareMathSymbol{\upi}{0}{UPM}{"19}
      \DeclareMathSymbol{\umu}{0}{UPM}{"16}
      \DeclareMathSymbol{\upartial}{0}{UPM}{"40}
      \DeclareMathSymbol{\leqslant}{3}{AMSa}{"36}
      \DeclareMathSymbol{\geqslant}{3}{AMSa}{"3E}
    \fi
  \fi
\fi 

\ifCUPmtlplainloaded \else
  \ifAMStwofonts \else 
    \def\upi{\pi}
    \def\umu{\mu}
    \def\upartial{\partial}
  \fi
\fi

\title{Orbital dynamics of three-dimensional bars: \\I. The backbone
of 3D bars.  A fiducial case} 
\author[Ch.~Skokos et al.]
{Ch.~Skokos,$^1$ P.A.~Patsis,$^1$ E.~Athanassoula$^2$\\ $^1$Research
Center of Astronomy, Academy of Athens, Anagnostopoulou 14, GR-10673
Athens, Greece\\ $^2$Observatoire de Marseille, 2 Place Le Verrier,
F-13248 Marseille Cedex 4, France} \date{Accepted ....  Received ....;
in original form ....}
\pagerange{\pageref{firstpage}--\pageref{lastpage}}
\pubyear{2001}

\begin{document}

\maketitle

\label{firstpage}

\begin{abstract}
  In this series of papers we investigate the orbital structure of 3D
  models representing barred galaxies. In the present introductory
  paper we use a fiducial case to describe all families of periodic
  orbits that may play a role in the morphology of three-dimensional
  bars. We show that, in a 3D bar, the backbone of the orbital
  structure is not just the x1 family, as in 2D models, but a tree of
  2D and 3D families bifurcating from x1. Besides the main tree we
  have also found another group of families  of lesser
  importance around the radial 3:1 resonance. The families of this group
  bifurcate from x1 and influence the dynamics of the system only locally.
  We also find that 3D orbits elongated along the
  bar minor axis can be formed by bifurcations of the planar x2
  family. They can support 3D bar-like structures along the minor axis
  of the main bar. Banana-like orbits around the stable Lagrangian
  points build a forest of 2D and 3D families as well. The importance
  of the 3D x1-tree families at the outer parts of the bar depends
  critically on whether they are introduced in the system as bifurcations in
  $z$ or in $\dot{z}$.  
\end{abstract}


\section{Introduction}
A thorough understanding of the orbital structure in a barred galaxy potential
can provide useful insight to the stellar dynamics of barred galaxies and
therefore to the dynamical evolution of these objects, as reviewed e.g. by
Athanassoula (1984), Contopoulos \& Grosb{\o}l (1989), Sellwood \& Wilkinson
(1993) and Pfenniger (1996).  Stable periodic orbits trap around them regular
orbits and thus constitute the backbone of galaxy structure (Athanassoula,
Bienayme, Martinet et al. 1983). Thus the appearance of a given morphological
feature can often be associated with the properties of one of the main
families of periodic orbits.  In the '90s, starting with Athanassoula (1992a;
1992b), many papers have pointed out that the gaseous response to steady
barred potentials is, to a large degree, determined by the morphology of the
periodic orbits in the corresponding stellar case. Thus, orbital and gaseous
dynamics are linked. This has provided added incentive for studies of the
morphology and the stability of periodic orbits in Hamiltonian systems
representing disc galaxies.

Orbital theory has often provided useful information on the structure
of galactic bars. Thus it is now understood that a bar is basically
due to regular orbits trapped around the so called `x1' periodic
orbits, which are elongated along the bar major axis \cite{cg89}. Such
orbits do not extend beyond the corotation resonance, and in many
cases no suitable elongated orbits can be found beyond the 4:1
resonance. This led orbital theory to predict that bars should end at
or before corotation \cite{gco80}. Orbital theory was also able to
predict - at the right distance from the center - the loops of the
near-infrared isophotes (see the case of NGC 4314 in Patsis,
Athanassoula \& Quillen 1997). Yet not all important morphological
features have been so far explained with the help of periodic
orbits. Thus orbital theory has difficulties to explain the boxy
isophotes surrounding the bars of, mainly, early-type barred galaxies
(Athanassoula 1996, Patsis et al. 1997). Another point still under
discussion is the morphology of the peanut-shaped bulges observed in
edge-on disc galaxies. They are considered by many authors as
revealing the presence of a bar, and to be associated with the 2:1
vertical resonance. It is not clear, however, which families can make
this vertical structure. Could a bar without a vertical 2:1 resonance
be boxy or peanut-shaped when viewed edge-on?  Could we have stellar
rings out of the equatorial plane at the ILR region?  Furthermore, the
detailed dynamics of the corotation region and the differences in the
vertical structure between fast and slow bars remain open issues.

In this series of papers we use orbital theory to address the above
questions.  This is a first step towards understanding both the
orbital behavior in $N$-body models and the responses of gaseous discs
to potentials derived from near-infrared observations.  The
differences between our model and the well studied corresponding 2D
case of the Ferrers bar \cite{ath92a} reflect the changes due to the
inclusion of the third dimension.  In separate papers we address the
question of the morphology of the peanut shaped-bulges (Patsis, Skokos
\& Athanassoula, 2002a - paper III) and of the boxy isophotes of bars
seen face-on (Patsis, Skokos \& Athanassoula, 2002b - paper IV).

Our first goal is to make a thorough study of the orbital structure
in 3D barred potentials, to classify the important families, and to follow
their morphological evolution as a function of the Jacobi integral. We
start with a fiducial case. Many of the families we find in this model
have been previously mentioned (e.g Heisler, Merritt \& Schwarzschild
1982; Pfenniger 1984, 1985b; Martinet \& de Zeeuw 1988; Hasan,
Pfenniger \& Norman 1993).  However, other, equally important
families, have not yet been studied.

Studying the orbital stability in a Hamiltonian system approximating the
dynamics of a barred galaxy we get the periodic orbits that could be used as
building blocks for a density model. The general rule is to look for stable
periodic orbits, since they trap around them the regular orbits. {\it Not all}
of them, however, are equally important.  The isodensities of the model we use
show us the topological limits within which we should look for the significant
orbits.  Stable representatives of families of periodic orbits which do not
support the imposed morphology, i.e. that of a bar embedded in an axisymmetric
disc with a central bulge, should be considered as less important.  As an
example, in 2D models, let us mention the case of the retrograde family x4,
which is stable over a very large interval of energies (Athanassoula et al.
1983). This family, however, should get a minimum weight when one tries to
construct a self consistent model using a Schwarzschild (1979) or a
Contopoulos \& Grosb{\o}l (1988) method. In a model of a 3D disc galaxy,
besides the counter-rotating x4 family on the equatorial plane, one has to
filter out also stable orbits with large $\overline{|z|}$, i.e. orbits with
large mean vertical deviations, since these orbits do not contribute much to
the density of the barred galaxy.

This paper is organized as follows: In section 2 we review briefly the parts
of orbital theory that are necessary for understanding this paper. In
particular we explain the use of characteristic and stability diagrams in
following the dynamical evolution of a family of periodic orbits. We describe
also the various types of instabilities encountered in 3D Hamiltonian systems
and we introduce the nomenclature of the main families. The latter is
necessary since a number of the families presented here have not been
previously discussed and thus need to be incorporated in a unique nomenclature
scheme.  In section 3 we introduce our 3D model and the orbital structure in a
2D counterpart. In section 4 we present the main families x1, x2 and x3 and
their bifurcations. In section 5 we describe the orbits around L$_4$ (or
L$_5$) and around L$_1$ (or L$_2$), as well as families outside corotation. We
conclude in section 6.

\section{A short introduction to periodic orbits in the present context}
\subsection{Periodic orbits and their stability}
In this section we will briefly review some parts of orbital theory 
which are necessary for the understanding of this paper. A clear, easily 
readable introduction to the subject has been given by Sellwood 
\& Wilkinson (1993). We also refer the reader to 
the pioneering works of Pfenniger (1984, 1985b) and 
Contopoulos \& Magnenat (1985). 

We study the stability of simple-periodic
orbits in a barred potential in cartesian coordinates. The
3D bar is rotating around its short $z$ axis. The $x$ axis is the intermediate
and the $y$ axis the long one. The system is rotating with an angular speed
$\Omega_{b}$ and the Hamiltonian governing the motion of a test-particle can
be written in the form:

\begin{equation}
H= \frac{1}{2}(p_{x}^{2} + p_{y}^{2} + p_{z}^{2}) +
    V(x,y,z) - \Omega_{b}(x p_{y} - y p_{x})    ,
\end{equation}
where $p_{x},~ p_{y},$ and $p_{z}$ are the canonically conjugate momenta. We
will hereafter denote the numerical value of the Hamiltonian by $E_j$ and
refer to it as the Jacobi constant or, more loosely, as the `energy'.
The corresponding equations of motion are:
\begin{eqnarray}
\dot{x}=p_{x}+\Omega_{b}y, & \dot{y}=p_{y}-\Omega_{b}x, &
\dot{z}=p_{z} \nonumber \\
\dot{p_{x}}= -\frac{\partial V}{\partial x} + \Omega_{b}p_{y}, &
\displaystyle{\dot{p_{y}}=-\frac{\partial V}{\partial y} - \Omega_{b}p_{x},} & 
\dot{p_{z}}=-\frac{\partial V}{\partial z}
\end{eqnarray}

The space of section in the case of a 3D system is 4D. The equations
of motion are solved for a given value of the Hamiltonian, starting
with initial conditions $(x_{0},\dot{x}_{0},z_{0},\dot{z}_{0})$ in the
plane $y$=0, for $\dot{y} > 0$. The next intersection with the $y$=0
plane with $\dot{y} > 0$ is found and the exact initial conditions for
the periodic orbit are calculated using a Newton iterative method. A
periodic orbit is found when the initial and final coordinates
coincide with an accuracy at least 10$^{-10}$.  The integration scheme
used was a fourth order Runge-Kutta scheme.

The estimation of the linear stability of a periodic orbit is based on
the theory of variational equations.  We first consider small
deviations from its initial conditions, and then integrate the orbit
again to the next upward intersection. In this way a transformation
$T: \mathbf{R}^{4} \to \mathbf{R}^{4}$ is established, which relates
the initial with the final point. The relation of the final deviations
of this neighboring orbit from the periodic one, with the initially
introduced deviations can be written in vector form as:
$\vec{\xi}=M\,\vec{\xi_{0}}$. Here $\vec{\xi}$ is the final deviation,
$\vec{\xi_{0}}$ is the initial deviation and $M$ is a $4 \times 4$
matrix, called the monodromy matrix.  It can be shown that the
characteristic equation is written in the form $\lambda^{4} + \alpha
\lambda^{3} + \beta \lambda^{2} + \alpha \lambda + 1 = 0$. Its
solutions $(\lambda_i, i=1,2,3,4)$ obey the relations
$\lambda_{1}\,\lambda_{2}=1$ and $\lambda_{3}\,\lambda_{4}=1$ and for
each pair we can write:
\begin{equation}
\lambda_{i}, 1/\lambda_{i} = \frac{1}{2} 
[-b_{i}\pm(b_{i}^{2} -4)^{\frac{1}{2}}],
\end{equation}
where $\displaystyle b_{i} = 1/2\,( \alpha \pm \Delta^{1/2})$ and $\Delta =
\alpha^{2} - 4 (\beta - 2)$.

The quantities $b_{1}$ and $b_{2}$ are called the stability
indices. If $\Delta > 0$, $|b_{1}|<2$ and $|b_{2}|<2$, the 4
eigenvalues are on the unit circle and the periodic orbit is called
`stable'. If $\Delta > 0$, and $|b_{1}|>2$, $|b_{2}|<2$, or
$|b_{2}|>2$, $|b_{1}|<2$, two eigenvalues are on the real axis and two
on the unit circle, and the periodic orbit is called `simple
unstable'.  If $\Delta > 0$, $|b_{1}|>2$, and $|b_{2}|>2$, all four
eigenvalues are on the real axis, and the periodic orbit is called
`double unstable'.  Finally, $\Delta < 0$ means that all four
eigenvalues are complex numbers but {\em off} the unit circle. The
orbit is characterized then as `complex unstable' (Contopoulos \&
Magnenat 1985, Heggie 1985, Pfenniger 1985a,b).  We use the symbols
$S,\: U,\: D,\: \Delta$ to refer to $stable$, $simple~unstable$,
$double~unstable$ and $complex~unstable$ periodic orbits respectively.
For a general discussion of the kinds of instability encountered in
Hamiltonian systems of {\sf N} degrees of freedom the reader may refer
to Skokos (2001).

The method described above has been initially presented by Broucke (1969) and
Hadjidemetriou (1975), and has been used in studies of the stability
of periodic orbits in systems of three degrees of freedom.  The reader
is referred to Pfenniger (1984) and Contopoulos \& Magnenat (1985) for an
extended description.

A diagram that describes the stability of a family of periodic orbits
in a given potential when one of the parameters of the system varies
(e.g. the numerical value of the Hamiltonian $E_j$), while all other
parameters remain constant, is called a `stability diagram'
(Contopoulos \& Barbanis 1985, Contopoulos \& Magnenat 1985). With the
help of such a diagram one is able to follow the evolution of the
stability indices $b_{1}$ and $b_{2}$, and the transitions from
stability to instability or from one to another kind of
instability. We will loosely refer to the $b=2$ and $b=-2$ lines on a
stability diagram as the $b=2$ and $b=-2$ axes. The S\ar U
transitions, when one of the stability indices has an intersection
with the $b=-2$ axis, or tangencies of the stability curves with the
$b=-2$ axis, are of special importance for the dynamics of a
system. In this case a new stable family is generated by bifurcation
of the initial one and has the same multiplicity as the parent
family. That means that the periodic orbits of the bifurcating family
have, before closing, as many intersections with the plane $y$=0, for
$\dot{y} > 0$, as the orbits of the parent family. The new family may
play an important role in the dynamics of the system. S \ar U
transitions after an intersection of a stability curve with the $b=2$
axis, or tangencies of a stability curve with the $b=2$ axis, also
generate a stable family but are accompanied by period doubling. This
means that the number of intersections with the plane $y$=0 (always
with $\dot{y} > 0$), needed for the periodic orbits to close, is
double the corresponding number of the parent family. Since the most
important families we examine here are simple-periodic, i.e. of
multiplicity 1, intersections or tangencies of their stability indices
with the $b=2$ axis introduce in the system families of orbits with
multiplicity 2. U\ar D and D\ar \D transitions do not bring new stable
families in the system and thus in principle are only of theoretical
interest. As we will see, however, the evolution of a family which is
found to be initially unstable may be very important for the dynamics
of our model. The family could simply become stable in another energy
interval, or it may play a major role in a collision of bifurcations,
an inverse bifurcation or other dynamical phenomena
\cite{gco86}. Finally in the case S\ar \D we have in general no
bifurcating families of periodic orbits.

Another very useful diagram is the `characteristic' diagram
\cite{cme}. It gives the $x$ coordinate of the initial conditions of
the periodic orbits of a family as a function of their Jacobi constant
$E_j$.  In the case of orbits lying on the equatorial plane and
starting perpendicular to the $x$ axis, we need only one initial
condition, $x$, in order to specify a periodic orbit on the
characteristic diagram. Thus, for such orbits this diagram gives the
complete information about the interrelations of the initial
conditions in a tree of families of periodic orbits and their
bifurcations. However, even for orbits completely on the equatorial
plane, but not starting perpendicular to the $x$ axis we need to give
initial conditions as position--velocity pairs $(x,\dot{x})$ and the
characteristic diagram is three-dimensional $(E_j,x,\dot{x})$. In the
general case of orbits in a 3D system, one has a set of four initial
conditions and the characteristic diagram is five-dimensional. The
representation of such a diagram is difficult, but when necessary we
will give just the $(E_j,x)$ projection. $(E_j,x)$ diagrams that can
be compared with the corresponding 2D models will always be given. In
all characteristic diagrams the region to which the orbits are
confined is bounded by a curve known as the zero velocity curve (ZVC),
since the velocity on it becomes zero.

\subsection{The nomenclature of the main families}
Our orbital study is more extended than previous ones and thus brings
in new families of orbits which have not been studied so far. We were
thus brought to introduce a new nomenclature system, extension of the
Contopoulos \& Grosb{\o}l (1989) system, which covers all the new types 
of orbits.

For the main 2D families of simple periodic orbits the nomenclature in
the present paper follows the standard notation of Contopoulos \&
Grosb{\o}l (1989). We thus have the x1 family, where orbits are
elongated along the bar and which is the main family in the case of
barred potentials, families x2 and x3, whose orbits are elongated
perpendicular to the bar, and the retrograde family x4.  2D families
bifurcated from x1 at the 3:1 resonance region on the equatorial plane
are denoted by t1, t2, ..., for consistency with the names used in
Patsis et al. (1997).  2D families bifurcated at the 4:1 resonance
region on the equatorial plane are called q1, q2, q3,....  Planar
orbits related with the 1:1 radial resonance will be called o1,
o2.... They are encountered only in some models. The fiducial case
presented in the present paper is not one of them.

Further planar families  appear beyond the x1 family,
at the gaps of the even resonances 4:1, 6:1, 8:1
etc. They are given the names `f', `s',
`e'... respectively. These families, not directly related to the
morphological problems we address in this series, will be discussed elsewhere. 

We name the 3D families bifurcated from the basic family x1 at the
vertical resonances as x1v$n$, where $n$ denotes the order of their
appearance in our fiducial model A (see below section 4). This is a convenient
model to be used for our nomenclature, since there are families of 3D orbits
associated with {\it all} basic vertical resonances. So x1v1 is
the one bifurcated at the first S\ar U transition, which happens at
the vertical 2:1 resonance region, x1v2 is the one bifurcated at the
U\ar S transition (second stability transition of the model also at
the vertical 2:1 resonance region), x1v3 is the one bifurcated at the
S\ar U transition at the vertical 3:1 resonance and so on.  Further
bifurcations of these x1v$n$ families are indicated with an `.$n$' (for the
$n$-th bifurcation) attached to the name of the parent family;
i.e. the first bifurcation of x1v1 will be x1v1.1, the second x1v1.2,
etc. Further bifurcations of these families will be indicated by further 
`.$n$' attached to the name of the parent family. Thus x1v1.1.1 is the 
first bifurcation of x1v1.1. The naming system is thus extendable at will.

In general at each vertical resonance we have two bifurcating families
introduced in the system. The number of oscillations along the
rotation axis $z$ corresponds to the vertical resonance at which the
family is born. E.g. families x1v1 and x1v2, which are bifurcated at
the vertical 2:1 resonance region, have orbits with two oscillations
along the $z$ axis. This determines only partially their morphology, since
the bifurcating family can be introduced either in the $z$ or the
$\dot{z}$ coordinate of the initial conditions.  If we know the number
of oscillations of a family along each axis and also whether it is a
bifurcation in $z$ or $\dot{z}$, then we know its morphology. Families
with similar morphology are similar in their corresponding $(x,y)$,
$(x,z)$ and $(y,z)$ projections. In  the fiducial case, where
each vertical resonance is associated with two bifurcating families, the
families x1v(2{\bf n}-3) and x1v(2{\bf n}-2) are born at the {\bf n}:1
resonance.

We note, however, that the first vertical
bifurcation is not in every model the x1v1 family, as in the fiducial
case.  In other models (Skokos, Patsis \& Athanassoula, 2002 - paper
II) it can happen that the first 3D bifurcation of x1 is not related
with the 2:1 vertical resonance, but with a different one. In such a
model the first vertical bifurcation of x1 will have the same name as
the family of the fiducial model which has similar morphology. 
Equivalently, it will have the same name as the family of the fiducial
model which is introduced in the same n:1 resonance and in the same
($z$ or $\dot{z}$) coordinate. This
way we make sure that families with similar morphologies share the
same name in the various models. In addition, if for some reason we have more
than one vertical bifurcation of x1 associated with a vertical resonance, we
introduce appropriate primes in our nomenclature. E.g. in a model with two
vertical 4:1 resonances we will have the pairs of bifurcating families x1v5, 
x1v6 and x1v5$^{\prime}$, x1v6$^{\prime}$. By keeping the basic name of
the family similar for all families associated with the same vertical
resonance, we underline again the dependence of the name on the encountered
morphology. Nevertheless, the basic names are given in the fiducial model,
which thus becomes a reference case for all our work.

We use the same nomenclature not only for the bifurcations of the
basic family x1, but in general for the vertical bifurcations of every
2D family. Their name consists of the name of the parent family,
followed by `v$n$', where $n$ indicates its $n$-th vertical
bifurcation. Also the names of the bifurcations of the bifurcating
families are characterized by the addition of `.1', `.2'$\dots$
etc. at the end of the name of the 3D family, as described above for
the corresponding families associated with x1.

We will use the same system in order to name also radially bifurcating
families. In general a radial bifurcation will be named as `wr$n$', where `w'
the name of the parent family. E.g. the $n$-th radial bifurcation of
family f will be `fr$n$' (fr1, fr2...etc.).

Let us now turn to orbits related with the axis of rotation. 
The family on the axis of rotation is called `z-axis'
family \cite{mdz88}. Its two first bifurcations are introduced at the
first S\ar U transition and the first U\ar S transition respectively
and they are the `sao' and `uao' orbits of Heisler, Merritt \&
Schwarzschild (1982). This nomenclature, however, does not lend itself
to extension which can include what Poincar\'{e} (1899) called the
`deuxi\`{e}me genre' families (cf. Polymilis, Servizi \& Skokos, 1997)
which can play an important role in some Ferrers bars (paper II), so
we will not adopt it here for other families related with the z-axis
orbits. {\em In practice} `deuxi\`{e}me genre' 
orbits are found on the stability diagrams as bifurcations of the
parent family when this family is considered as being of higher
multiplicity, i.e. if its orbits are repeated many times. Thus, the
z-axis family, when its orbits are repeated twice, is called
z2. Bifurcations of the z2 family are called z2.1, z2.2 etc. The same
rule applies for the bifurcations of z3, i.e. for the bifurcations of
z-axis if this is described three times. We then have z3.1, z3.2 and
so on. These bifurcating families always come in pairs. A further
index (s or u) is attached to their names and is related with their
stability.

Around the Lagrangian points $L_{4,5}$ we have the long period
banana-like orbits, which form a tree of families, and the short
period orbits. For the latter we keep the Contopoulos \& Grosb{\o}l
(1989) notation (spo). For the banana-like orbits we use the notation
ban1, ban2,...ban$n$ in the 2D cases. Their 2D bifurcations are the
families ban$n$.1, ban$n$.2,... etc. and their 3D bifurcations the
families ban$n$v1, ban$n$v2,... etc. 3D banana-like orbits not related
with a 2D one are named banv$n$.

A 2D family found around the unstable Lagrangian points $L_{2,3}$ is called
$\ell_1$. 

Throughout the papers we give also the names used by other authors for families
that have been previously studied. However, since our study is more extended,
there are several families mentioned  for the first
time.

\section[]{The model}
\subsection{The 3D potential}
For our calculations we used a 3D potential, which consists of a
Miyamoto disc, a Plummer bulge and a Ferrers bar. Pfenniger and
collaborators have made extensive use of this potential for orbital
calculations (Pfenniger 1984, Pfenniger 1985a;b, Martinet \& Pfenniger
1987, Pfenniger 1987, Pfenniger 1990, Hasan, et al. 1993,
Olle \& Pfenniger 1998). Our work is, in many ways, more extended. We
make a much more extensive search for periodic families and we
furthermore follow their stability. The latter allows us to find a
number of `new' families, which show interesting morphological
characteristics. Furthermore, we vary the parameters of the model so
that we are able to make comparisons between fast and slow rotating
bars as well as between strong and weak bars (paper II). Finally, we focus our
work more on tracing the orbital behaviour that could support observed
morphological features and less on studying in depth qualitatively the
dynamical phenomena that take place in this kind of Hamiltonian
systems. 

Our general model consists of 3 components.  The disc is represented
by a Miyamoto disc (Miyamoto \& Nagai 1975), the potential of which
reads:

\begin{equation}
\label{potd}
\Phi _{D}=-\frac{GM_{D}}{\sqrt{x^{2}+y^{2}+(A+\sqrt{B^{2}+z^{2}})^{2}}},
\end{equation}
where \( M_{D} \) is the total mass of the disc, $A$ and $B$ are the
horizontal and vertical scale lengths, and $G$ is the gravitational
constant. The bulge is modeled by a Plummer sphere with potential:
\begin{equation}
\label{pots}
\Phi _{S}=-\frac{GM_{S}}{\sqrt{x^{2}+y^{2}+z^{2}+\epsilon _{s}^{2}}},
\end{equation}
where \( \epsilon _{s} \) is the scale length of the bulge and \(
M_{S} \) is its total mass. The third component of the potential is a
triaxial Ferrers bar, whose density \( \rho (x) \) is:
\begin{equation}
\label{densd}
\rho (x)=\left\{ \begin{array}{lcc}
\displaystyle{\frac{105M_{B}}{32\pi abc}(1-m^{2})^{2}} & {\mbox for} &
m \lid 1\\ 
 & & \\
\displaystyle{0} & {\mbox for}  & m>1
\end{array}\right. ,
\end{equation}
where
\begin{equation}
\label{semiaxis}
m^{2}=\frac{y^{2}}{a^{2}}+\frac{x^{2}}{b^{2}}+\frac{z^{2}}{c^{2}}\, \, ,\, \,
 \, 
 a>b>c,
\end{equation}
\( a \), \( b \), \( c \) are the semi-axes and \( M_{B} \) is the
mass of the bar component. The corresponding potential \( \Phi _{B} \)
and the forces are given in Pfenniger (1984)\footnote{We made use of
the offer of Olle \& Pfenniger (1998) for free access to the
electronic version of the potential and forces routines.}.  They are
in a closed form, well suited for numerical treatment.  For the
Miyamoto disc we use A=3 and B=1, and for the axes of the Ferrers bar
we set $a:b:c = 6:1.5:0.6$, as in Pfenniger (1984). We note that
these axial ratios are near the standard values given by 
Kormendy (1982). The masses of the three
components satisfy \( G(M_{D}+M_{S}+M_{B})=1 \).  The length unit is taken as
1~kpc, the time unit as 1~Myr and the mass unit as $ 2\times 10^{11}
M_{\odot}$. 
      
In Table~\ref{tab:models} we give the parameters of our model.
We give it the name A1, and it will be one of the models to be used in our
comparative study in paper II. 
\begin{table*}
\caption[]{The parameters of our fiducial model A1. G is the
  gravitational constant, M$_D$, M$_B$, M$_S$ are the masses of the
  disk, the bar and the bulge respectively, $\epsilon_s$ is the scale
  length of the bulge, \O$_{b}$ is the pattern speed of the bar,
  $E_j$(r-IILR) and $E_j$(v-ILR) are the values of the Jacobi constant for
  the radial and vertical 2:1 resonances and $R_c$ is the corotation
  radius.}
\label{tab:models}
\begin{center}
\begin{tabular}{ccccccccc}
  GM$_D$ & GM$_B$ & GM$_S$ & $\epsilon_s$ & \O$_{b}$ & $E_j$(r-IILR)
 & $E_j$(v-ILR) & $R_c$ \\ 
\hline
  0.82 &  0.1  & 0.08 & 0.4 &  0.054 & -0.44 & -0.36& 6.13 \\

\hline
\end{tabular}
\end{center}
\end{table*}

\subsection{The 2D Ferrers bar}
The general orbital structure in potentials including a 2D Ferrers bar
can be found in Athanassoula (1992a). The dynamics are dominated by
the presence of the x1 family, which is in general stable. It is
characterized by the presence of a narrow instability zone at the 3:1
resonance and a gap at the 4:1 region, which is generally of type 2
\cite{cg89}.  The S\ar U\ar S transition at the 3:1 region introduces
in the system a couple of simple periodic families of orbits, the
importance of which remains local. Beyond the type 2 gap and above the
local maximum of the characteristic of x1 at the 4:1 resonance (Fig. 2
in Contopoulos \& Grosb{\o}l 1989) one can find a large number of
families squeezed close to the zero velocity curve. Finally the
families x2 and x3 generally exist for a large energy range and their
characteristics form a single bubble.  As it is known, x2 is generally
stable and x3 unstable.

In the next sections we describe the orbital behaviour in a 3D case
where both radial and vertical 2:1 resonances exist. We will thus find
the differences introduced in the morphology and stability of the
families of periodic orbits by the inclusion of the third
dimension. We will also examine how the 3D families of periodic orbits
support the bar.

\section{The $\bmath{\lowercase{\mbox{{x1}}}}$-family and its bifurcations}
\subsection{A general description}
Contrary to the 2D models, where a single family, the x1 family,
provides the building blocks for the bar, in 3D models we have a {\em
tree of families} consisting of 2D and 3D families related to the
planar x1 orbits. In Table~\ref{tab:x1tree} we summarize the
properties of these families.  We list their name, the value of the
energy at which they are born ($E_j^*$), the $E_j$ intervals where
they are stable and we indicate whether they are two-dimensional (2D)
or three-dimensional (3D). Their interconnections and their role will
be described in the following paragraphs.

\begin{table*}
\caption[]{The families of the x1-tree. The successive columns give
  the name of the families, the value of the energy at which they are
  introduced ($E_j^*$), the intervals of $E_j$ at which they are
  stable and also if they are 2D or 3D. The `bow'-region is explained
  in the text, while `...' after an energy value indicate that a
  family continues to be stable, but reaches distances far away from
  the $z=0$ plane.}
\label{tab:x1tree}
\begin{center}
\begin{tabular}{cccc} 
\hline
family & $E_j^*$ & stable intervals in $E_j$ & 2D / 3D \\ \hline
x1 & $-0.495$ & $-0.495 < E_j < -0.360$    & 2D \\
 & & $-0.343 < E_j < -0.293$ &  \\
 & & $-0.278 < E_j < -0.244$ &  \\
 & & `bow'-region $-0.222 < E_j < -0.214$  &  \\
 & & $-0.211 < E_j < -0.205$ &  \\
 & & $-0.192 < E_j < -0.191$ &  \\
 & & $-0.186 < E_j < -0.185$ &  \\
 & & $-0.175 < E_j < -0.173$ &  \\ \hline
x1v1 & $-0.360$ & $-0.360 < E_j < -0.336$ & 3D \\
 & & $-0.253 < E_j < -0.147 \,\ldots $  &  \\
x1v2 & $-0.343$ & always unstable & 3D \\
x1v3 & $-0.293$ & $-0.293 < E_j < -0.221$ & 3D \\
x1v4 & $-0.278$ & $-0.224 < E_j < -0.149$ & 3D \\
x1v5 & $-0.213$ & $-0.213 < E_j < -0.172$ & 3D \\
x1v6 & $-0.211$ & always unstable & 3D \\
x1v7 & $-0.205$ & $-0.205 < E_j < -0.183$ & 3D \\
 & & $-0.174 < E_j < -0.170$ &  \\
x1v8 & $-0.192$ & always unstable & 3D \\
x1v9 & $-0.185$ & $-0.185 < E_j < -0.182$ & 3D \\ \hline
\end{tabular}
\end{center}
\end{table*}

There are also several 2D families, which are radial bifurcations of
x1 and thus part of the x1-tree, but play a less important role in
the morphology of the models. They are described in a separate table
(Table~\ref{tab:x1r}). The `t' families are related with the 3:1 and
the `q' with the 4:1 radial resonance region.
\begin{table*}
\caption[]{Radial bifurcations of the x1 family.  `t' families are
related with the 3:1  and `q' families with the 4:1 radial
resonance region. Columns are as in Table~\ref{tab:x1tree}. $E_j$
values are given in general with three digits, except from the cases
where narrow $E_j$ ranges of existence need more accuracy. We note
that the t3 family exists also for lower energies than its $E_j^*$ (see
section 4.4).}
\label{tab:x1r}
\begin{center}
\begin{tabular}{cccc} \hline
family & $E_j^*$ & stable intervals in $E_j$& 2D / 3D \\ \hline
t1 & $-0.244$ & $-0.244 < E_j < -0.218$    & 2D \\
t2 & $-0.214$ & $-0.214 < E_j < -0.209$, & 2D \\
 & & $-0.204 < E_j < -0.203 $  &  \\
t3 & $-0.205$ & $-0.2065 < E_j < -0.2005 $ & 2D \\
q1 & $-0.191$ & always unstable & 2D \\
q2 & $-0.1857$ & $-0.1860 < E_j < -0.1857 $ & 2D \\
q3 & $-0.183$ & $-0.1818 < E_j < -0.1808 $ & 2D \\ \hline
\end{tabular}   
\end{center}
\end{table*}

Besides the orbits related to the x1 family, we find the x2 and x3
families and their 3D relatives as well. They exist for the same
energy intervals as the families of the x1-tree, but their projections
on the equatorial plane are elongated along the minor axis of the
bar. They are described below.

\subsection{Families x1, x2 and x3}
\begin{figure}
\rotate[r]{ 
\epsfxsize=6.0cm \epsfbox{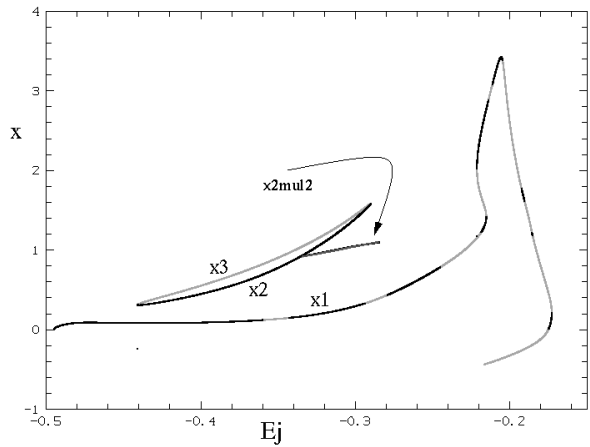}
}
\caption[]{x1, x2-x3 and the x2mul2 $(E_j,x)$-characteristics. The
curve corresponding to x2mul2 is the projection of its characteristic
on the $(E_j,x)$ plane. Stable regions are drawn in black and unstable
ones in light gray.}
\label{x1char}
\end{figure}

\begin{figure*}
\epsfxsize=14.0cm \epsfbox{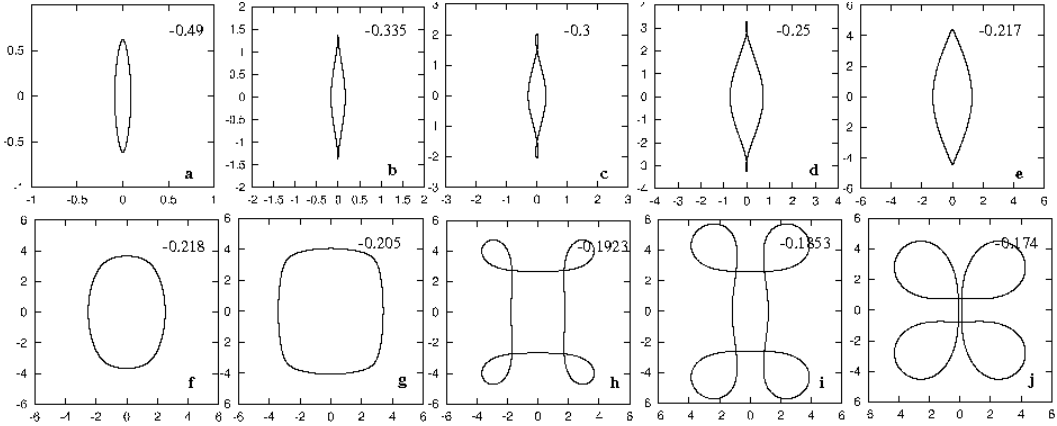}
\caption[]{x1 stable orbits in model A1. The numbers at the upper right
  corners of the panels indicate their $E_j$ values.}
\label{x1pan}
\end{figure*}

The characteristics of the x1 and the x2-x3 families in model A1
(Fig.~\ref{x1char}) have the typical geometry of the characteristics
of 2D Ferrers bars \cite{ath92a}. Due to the vertical instabilities,
however, x1 becomes unstable over several $E_j$ intervals, and not
only at the radial 3:1 resonance region, as in the 2D case. In
Fig.~\ref{x1char} and in all characteristic diagrams hereafter we draw
the unstable regions in light-grey. We observe that the decreasing
part of the x1 curve, below the local maximum at the radial 4:1
resonance region ($E_j \approx -0.21$), is almost everywhere light-grey,
indicating that the 
family is unstable there. The curve at about $E_j \approx -0.17 $
turns back towards lower energies, remaining after that continuously
unstable.
The morphological evolution of the x1 orbits is the one expected from
the 2D case and is given in Fig.~\ref{x1pan}. The numbers at the upper
right corners of the individual frames correspond to the $E_j$ value
of the orbit. The orbits are chosen along the characteristic curve
starting from the lower values of the Jacobi constant; the orbits in
Fig.~\ref{x1pan}h,i,j belong to the decreasing branch.  Except for the
instability zones related to the 3:1 resonance all other unstable
parts of x1 appear only in the 3D case. As mentioned in section 2.1,
the families introduced at the instability strips by bifurcation
inherit the kind of stability of the parent family, i.e. of x1. Thus,
the instability gaps on the x1 characteristic are covered 
by the stable orbits of the families born after the corresponding S\ar
U transitions. So for almost every energy $E_j$ there exists a stable orbit of
the x1-tree. As we mentioned in section 2, the 3D bifurcated
families are in general characterized by four initial conditions
$(x_{0},\dot{x}_{0},z_{0},\dot{z}_{0})$ so that a $(E_j, x_{0})$
characteristic diagram cannot provide all the essential
information. For this reason we prefer to follow the dynamical
evolution of the orbits using stability diagrams.
These diagrams frequently become complicated, but they have the big
advantage of giving in a straightforward way the interconnections of
the various families, thus becoming a very useful tool in the hunting
of periodic orbits.

\begin{figure}
\rotate[r]{  
\hspace{-0.5cm} 
\epsfxsize=6.2cm \epsfbox{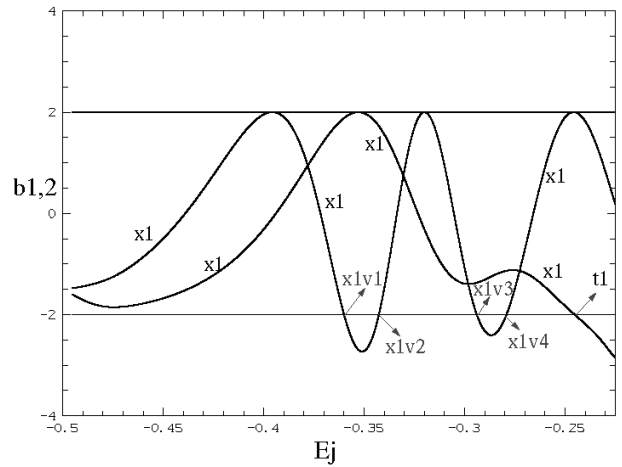}
}
\caption[]{First part of the x1 stability diagram. Arrows denote the
  bifurcation of families and the direction in which these evolve.}
\label{x1-I}
\end{figure}

\begin{figure}
\rotate[r]{ 
\epsfxsize=6.2cm \epsfbox{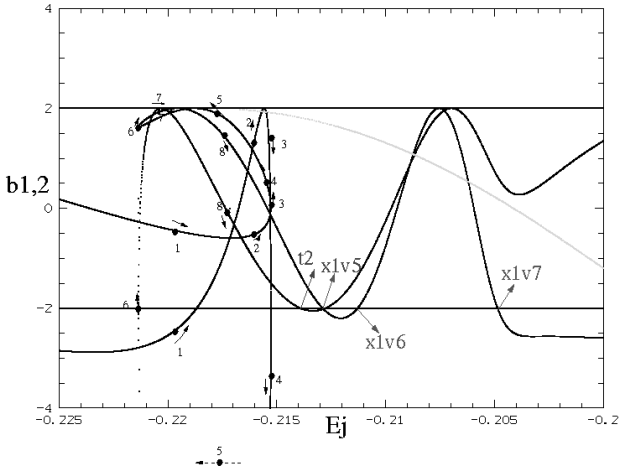}
}
\caption[]{Second part of the x1 stability diagram. The evolution of
  the stability curves in the x1 `bow' is indicated 
  with numbers from 1 to 8 and arrows. In the diagram we indicate also the x1
  bifurcations to the right of the `bow'. They are the families t2, 
  x1v5, x1v6 and x1v7.}
\label{x1-fiogkos}
\end{figure}

\begin{figure}
\rotate[r]{ 
\epsfxsize=6.2cm \epsfbox{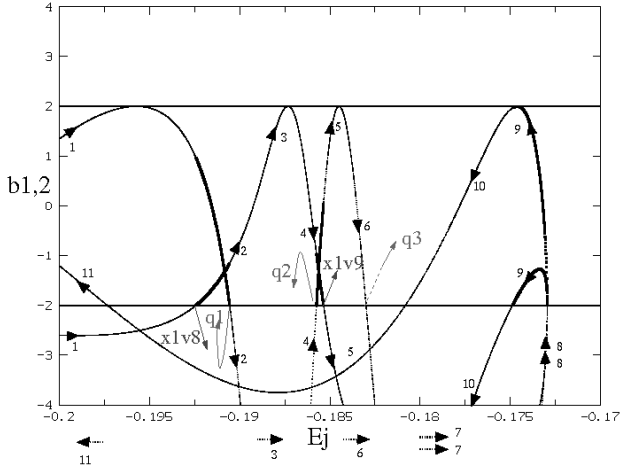}
}
\caption[]{Stability diagram for x1 and bifurcating families
corresponding to the decreasing part of the x1 characteristic. The
evolution of the stability curves is indicated with numbers from 1 to
11, and thick arrows that point to the direction of the evolution. The
bifurcating families and their direction of evolution are denoted with
thin arrows.}
\label{x1-III}
\end{figure}

The evolution of the stability indices $b_1$ and $b_2$ for x1 are given in
Figs.~\ref{x1-I}, ~\ref{x1-fiogkos}, and ~\ref{x1-III} for successive energy
intervals. The arrows denote bifurcated families at the bifurcating
points and show the direction of the stability index associated
with the S\ar U or U\ar S transition. We observe that the variation of the
index which in Fig.~\ref{x1-I} has the larger values for $E_j < -0.38$
brings in the system the 3D families x1v1, x1v2 etc., while the
variation of the other index brings in the families associated with
the radial instabilities. The latter remain on the equatorial
plane. The variation of their stability indices will in turn bring new
families due to vertical and radial instabilities.

The feature depicted in Fig.~\ref{x1-fiogkos} is typical of the
stability diagrams of many of our models. We call this kind of
evolution of the stability indices a `bow'. The $b_1$ and $b_2$ curves
do not break anywhere, but they evolve in a continuous, rather
complicated way, changing direction twice. This `bow' area corresponds
to the bend, or elbow, in the characteristic at about $E_j \approx -0.227$
(Fig.~\ref{x1char}), and the complicated evolution of the stability
indices happens as we move  towards lower $E_j$ values along
the  characteristic curve of x1 at this area. In
Fig.~\ref{x1-fiogkos} one can follow the evolution of $b_1$ and $b_2$
by following the evolution of both the numbers and the nearby
arrows. The lowest value of the stability index at `5', not included
in the Figure (indicated only with a dashed arrow outside of figure
frame), is $\approx -55$.

A significant change in the way the 3D bifurcations of x1 are introduced in
the system happens at the instability zone found just beyond the local maximum
of the $(E_j,x)$ characteristic close to the radial 4:1 resonance. As we see
in Fig.~\ref{x1-I}, \ref{x1-fiogkos} and \ref{x1-III}, the 3D families are
bifurcated at S\ar U and U \ar S transitions, where the corresponding
stability index intersects the $b=-2$ axis.  Moving on the characteristic
towards corotation, before reaching the decreasing branch, a bifurcating
family at an S\ar U transition is a stable 3D family with initial conditions
$(x,z,\dot{x},\dot{z})=(a,b,0,0)$, where $a,b \in \mathbf{R}$ and $a,b \neq
0$. On the other hand, the family bifurcated at the U\ar S transition, is
(initially) simple unstable and has initial conditions
$(x,z,\dot{x},\dot{z})=(c,0,0,d)$, with $c,d\in \mathbf{R}$ and $c,d \neq 0$.
This means that the family introduced in the system as stable is a bifurcation
at $z$, and the simple unstable family a bifurcation at $\dot{z}$.  For the
set of families associated with the vertical 5:1 resonance, on the decreasing
branch of the characteristic, this sense of bifurcation is reversed. Namely we
have the bifurcation in $z$ at the U\ar S transition (x1v8) and the
bifurcation in $\dot{z}$ at S\ar U (x1v7).

In Fig.~\ref{x1-III} we plot the last part of the stability diagram of
the x1 family, corresponding to energies higher than $-0.2$.
As can be seen from the characteristic diagram of Fig.~\ref{x1char},
this includes most of the decreasing part of the characteristic, the
bend at $E_j \approx -0.173$ and the part that goes towards lower
energies. This part (roughly for $-0.22 < E_j < -0.173$), has negative
$x$ values starting soon after the bend. Heavy arrows and numbers in
increasing order on and next to the stability curves in
Fig.~\ref{x1-III} indicate the evolution of the indices as we move
along this part of the characteristic. As we can see most parts are
unstable, the short stable parts being drawn with heavy lines.  After
the turning point, at $E_j \approx -0.173$, the upper curve, moving
now towards lower $E_j$ values, is stable until $E_j \approx -0.181$,
then has a part with values smaller than $-2$ and then reenters the
stability region for $E_j \approx -0.197$. The lower stability curve,
however, reaches absolutely large negative values. Thus, the family is
always unstable in the parts where $x < 0$. It is easy to understand
how the negative $x$ values are introduced by following the evolution
of the x1 orbit morphology as we move along the characteristic
(Fig.~\ref{x1pan}). As one moves along the decreasing part of the
characteristic (Fig.~\ref{x1pan}h \ar j), the four apocentra of the
orbits develop loops, whose size increases strongly as the energy
increases. Already for the orbit in Fig.~\ref{x1pan}j the loops have
become so large, that the sides of the orbit along the bar major axis
nearly touch. As we continue along the characteristic they will touch
and then cross, so that $x$ becomes negative.

\begin{figure*}
\rotate[r]{   
\epsfxsize=10cm \epsfbox{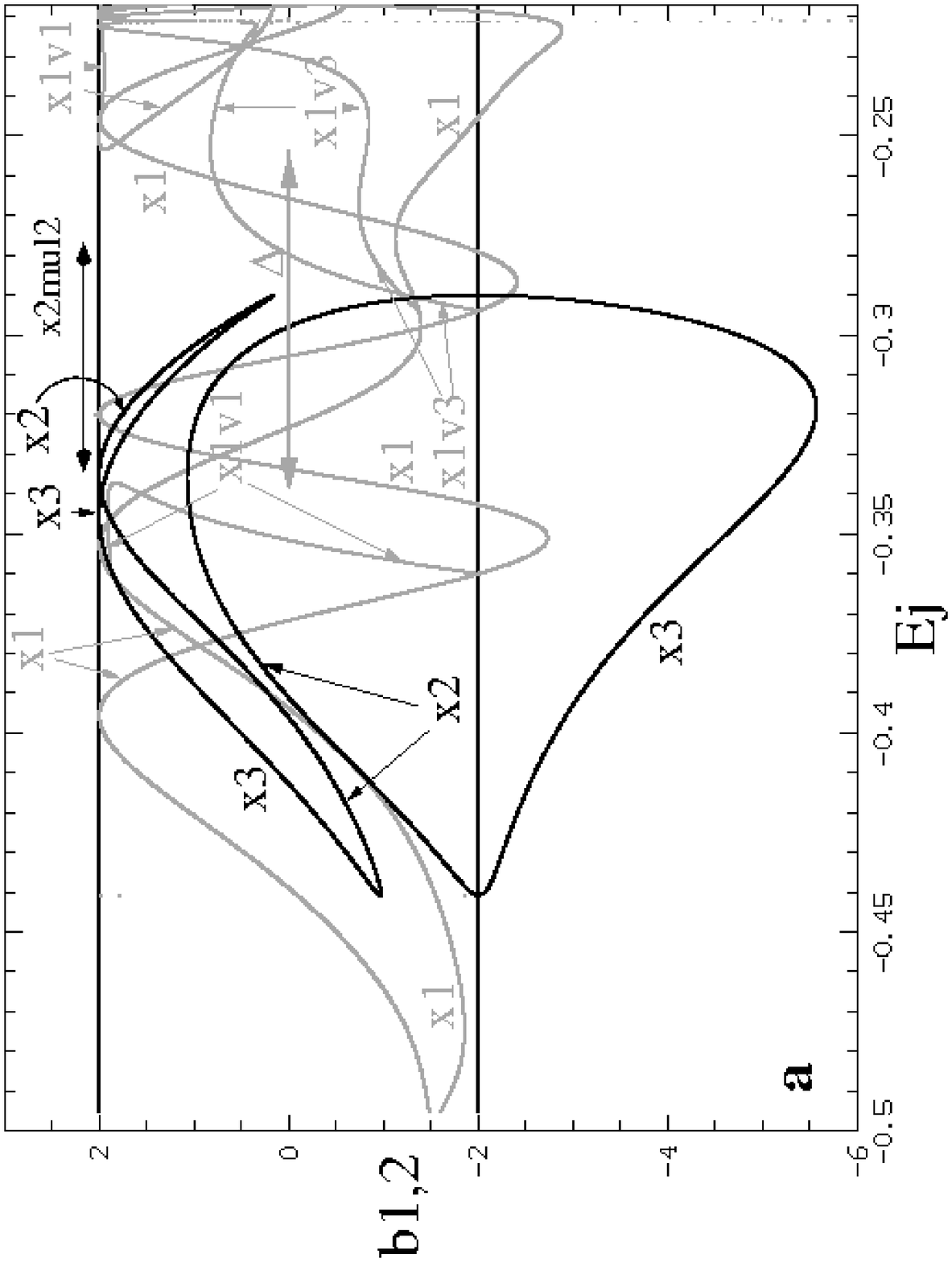}
\epsfxsize=10cm \epsfbox{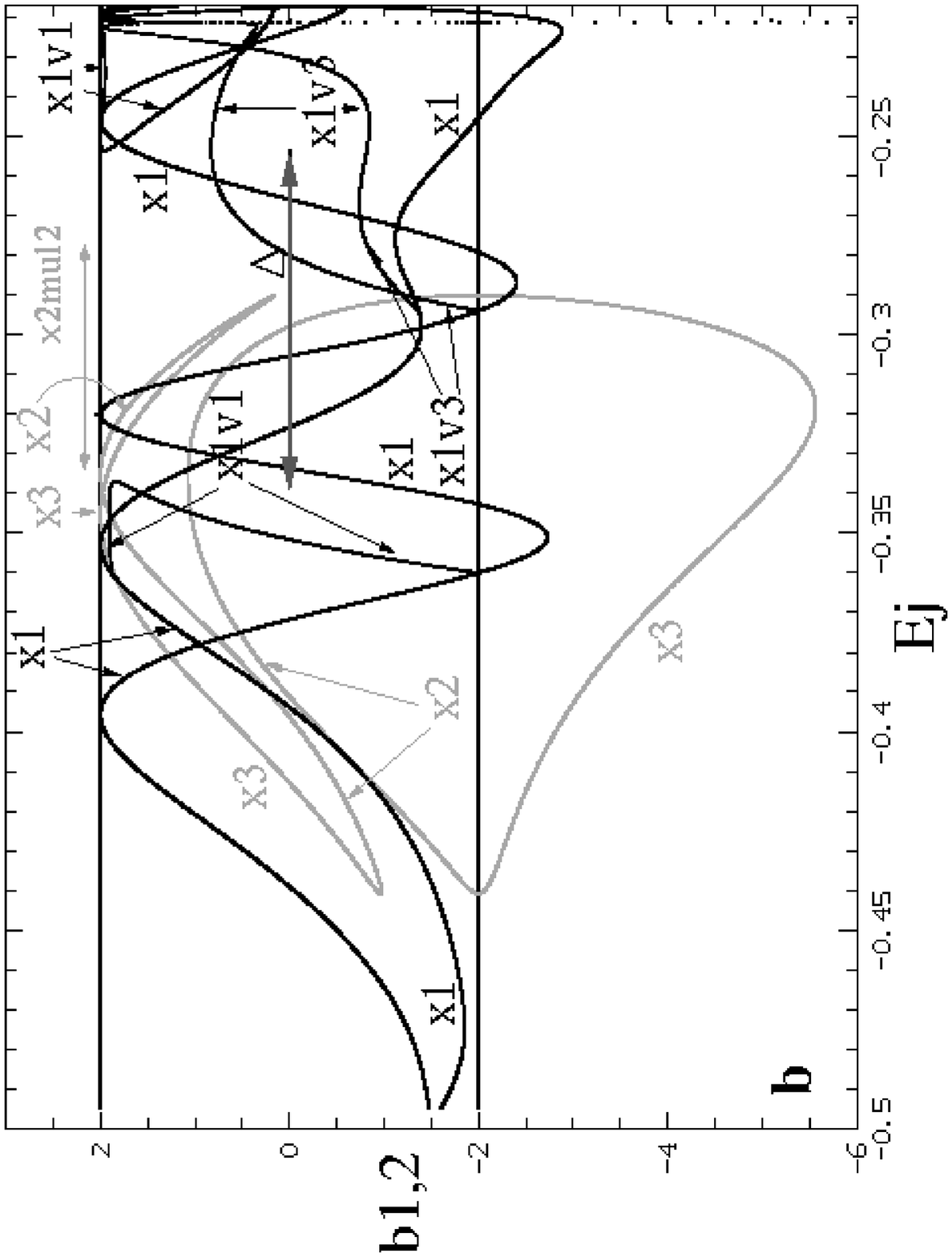}
}
\caption[]{Stability diagram for x1, x2 and x3 orbits. In order to
  follow the interconnections of the various families, this diagram is given in
  two parts. In (a) the stability indices of the families x2 and x3
  are emphasized and in (b) those of x1 and of the bifurcating
  families. A horizontal segment with double arrows in the upper part
  of the diagrams indicates the range of stability of the family
  x2mul2. The horizontal segment with double arrows, drawn black in
  the lower panel and indicated with $\Delta$, denotes the complex
  unstable part of x1v1.}
\label{x2-area}
\end{figure*}

\begin{figure}
\rotate[r]{   
\epsfxsize=3.0cm \epsfbox{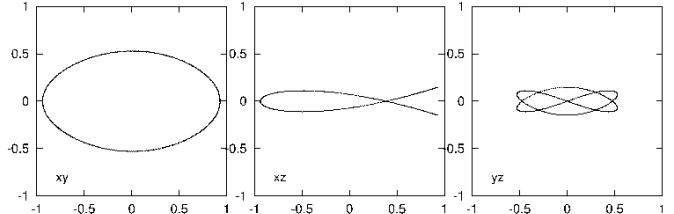}
}
\vspace{-0.5cm}
\caption[]{The 3D x2-like orbit x2mul2. This is a family of
 multiplicity 2.}
\label{x2-3d}
\end{figure}

Let us now present the evolution of the x2-x3 loop in the 3D case. As
seen in Fig.~\ref{x1char}, the situation with the x2-x3 characteristic
is exactly like in 2D. The stability indices form also loops, as the
$b_1$ and $b_2$ indices of x2 and x3 join each other in pairs
(Fig.~\ref{x2-area}).
In 2D models, the families x1 and x2 are in general the only simple
periodic stable families at the x2-x3 area. This is not necessarily the
case in 3D models. E.g. in this model, as we can see in
Fig.~\ref{x2-area}b, the 3D family x1v1 has been bifurcated as stable
just before the point $E_j =-0.36$, while close to $E_j =-0.29$ the
family x1v3 is introduced in the system. So the situation at the x2-x3
area is more complicated, since we have there {\em four} simple
periodic stable families. Since the x2-x3 stability indices form a
bubble they have no further intersections with the $b=-2$ axis and
there are no further bifurcations of other simple periodic x2-like
families.  Both families, however, have tangencies with the $b=2$
axis. At these points, as mentioned in the introduction, families of
the same kind of stability, but with double multiplicity, will be
bifurcated. The one bifurcated from the stable family x2 is
interesting. If we put its $x$ initial values on the characteristic
diagram (Fig.~\ref{x1char}), we obtain the extra branch emerging from
the x2-x3 loop, pointed with the curved arrow and characterized as
`x2mul2'. The energy range over which it is stable is indicated with a
double arrow above the $b=2$ axis in Fig.~\ref{x2-area}a. Its
morphology is given in Fig.~\ref{x2-3d}. The $(x,y)$ projection is
typical of an x2 orbit, the $(x,z)$ one is a fish-like figure
reflecting the double multiplicity of the family, while the $(y,z)$
projection offers a shape that could produce a tiny boxy structure in
the central region of the bar (note the scale on the axes). The $(x,z)$
projection can also offer a boxy structure, if one considers together
with every orbit its symmetric with respect to the $z$ axis. This,
however, is elongated along the {\em minor} axis of the bar as will be
discussed in paper III. The morphology of this family shows that the
model clearly can support in its face-on projection the presence of
stellar rings in the x2-x3 area. This, however, is a {\em thick} ring
structure extending outside the equatorial plane.

\subsection{The main 3D families}
The x1 S\ar U transition at about $E_j \approx -0.36$
(Fig.~\ref{x2-area}b), generates the 3D family of periodic orbits
x1v1. This family is related with the presence of the vertical 2:1
resonance. It has a stable part close to the bifurcating point, then
it has a complex unstable part after an S\ar \D transition, and
becomes again stable at about $E_j \approx -0.253$. We have found x1v1
as stable up to $E_j \approx -0.147$.

\begin{figure}
\rotate[r]{    
\epsfxsize=8.2cm \epsfbox{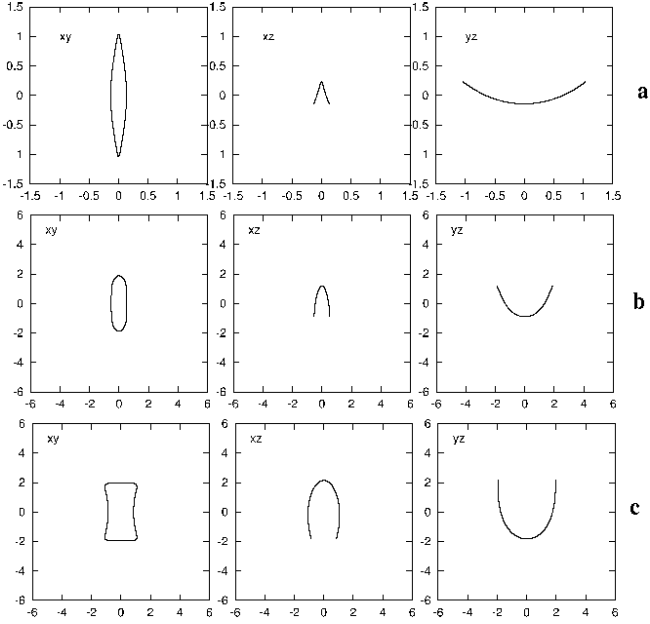}
}
\caption[]{Three stable orbits of the x1v1 family. Note that the upper
  panels have a different scale than the middle and  lower
  ones. Corotation in model A1 is at 6.13. The energies from top to
  bottom are: $E_j = -0.35$, $-0.25$ and $-0.20$ respectively.}
\label{x1-v1all}
\end{figure}

The morphological evolution of the family x1v1 is given in
Fig.~\ref{x1-v1all}. This family corresponds to the z2 family of Hasan et al.
(1993) and its orbits have
been associated with the appearance of the peanut shaped bulges by
Combes, Debbasch, Friedli et al. (1981). Indeed, due to the symmetry of the
potential with respect to the equatorial plane, one can find all 3D
families in pairs. Thus  for x1v1 we will have the smile
($\smile $) and frown ($\frown $) types of the $(y,z)$ edge-on
projections coexisting at a given energy, and the same holds for the 
$(x,z)$ projection. The $(x,y)$ projections of the 3D orbits follow in general
the 
morphology of the corresponding x1 orbit of the same energy. As we
said in the introduction, the importance of a family of stable
periodic orbits is limited as the individual orbits grow in $|z|$. The
x1v1 orbit for $E_j =-0.2$ in Fig.~\ref{x1-v1all}c exceeds both in its
$(x,z)$ and its $(y,z)$ projections the height of 2~kpc and this means
that it cannot contribute significantly to the density of the galactic
disc. Its spatial extent on the other hand indicates that this orbit
could be used to populate the bulge area.

The U$\rightarrow$S transition at $E_j \approx -0.343$ generates the
family x1v2 which we followed until $E_j\approx -0.173$. It remains
totally unstable and ends after a U $\rightarrow$ D $\rightarrow $ \D
sequence. It thus doesn't play any important role in the dynamics of
the system.

\begin{figure}
\rotate[r]{   
\epsfxsize=8.2cm \epsfbox{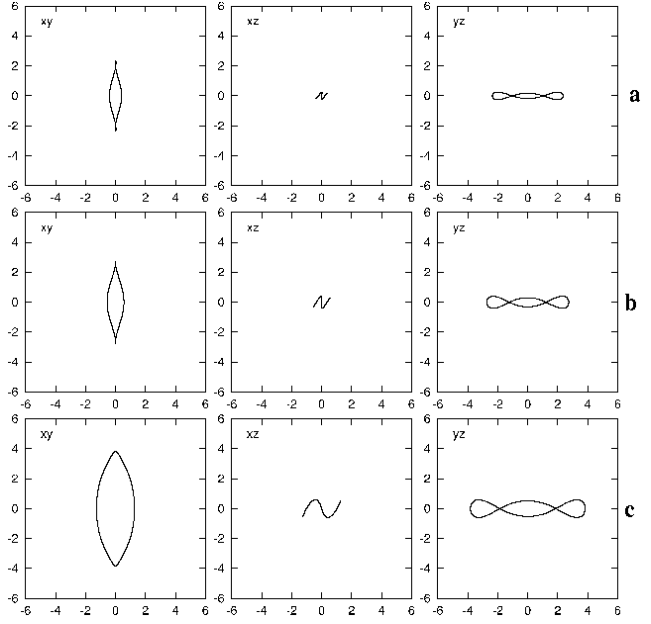}
}
\caption[]{Three stable orbits of the x1v3 family. Their $(x,z)$ and
$(y,z)$ projections have always low $|z|$ values. The energies from
top to bottom are: $E_j = -0.28$, $-0.26$ and $-0.22$ respectively.}
\label{x1-v3all}
\end{figure}

Family x1v3 (Fig.~\ref{x1-v3all}) is stable and its orbits keep low
$|z|$ values roughly in the interval $-0.293<E_j< -0.221$. It then  ends with
a S\ar \D transition. This family is similar to the z1 family of Hasan et al.
(1993). We note that both x1v1 and x1v3
provide useful orbits in the system before their S\ar \D
transition. This behaviour is also seen in the 3D thick spiral model
in Patsis \& Grosb{\o}l (1996). Complex instability helps introducing
abrupt drops in the density of given features of a model (in our case
the peanut), since it stops abruptly the existence of the family
responsible for their appearance without bringing new stable families
in the system. On the other hand, in cases where a stable family
donates its stability to a bifurcation we have a smooth morphological
evolution, which can give smooth density profiles in the
galaxies. Both x1v1 and x1v3 do not have any intersections or
tangencies with the $-2$ axis and for this reason they do not
bifurcate other families with the same multiplicity.

\begin{figure}
\rotate[r]{ 
\epsfxsize=8.0cm \epsfbox{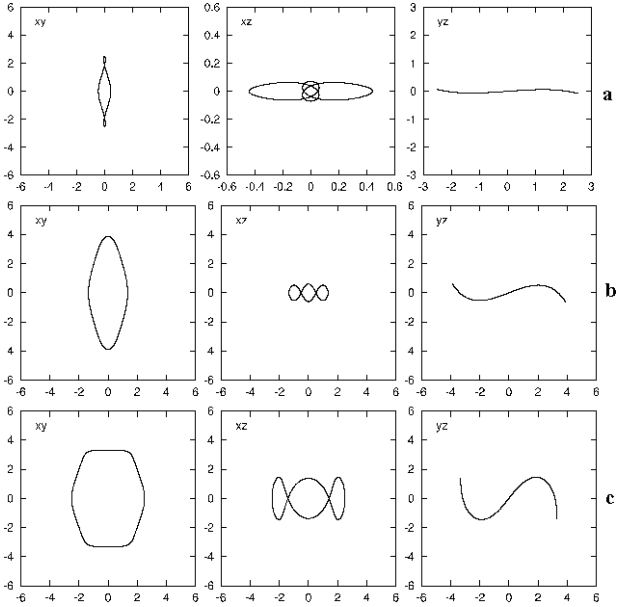}
}
\caption[]{Orbits of the family x1v4. Panels (a) show an unstable
orbit close to the bifurcating point at $E_j =-0.278$. Panels (b) and
(c) show stable orbits for $E_j =-0.22$ and $E_j =-0.206$
respectively. }
\label{orb-x1v4}
\end{figure}

The next bifurcated family is x1v4. This is bifurcated from x1 after a
U\ar S transition. We would thus have expected it  to be 
unimportant, since its parent family, x1, is unstable at the
bifurcating point. This is the typical behaviour in such cases and we
have seen it already happening for x1v2.  x1v4 is introduced at about
$E_j \approx -0.278$. 
%
One of the two stability indices,  let us call it $b_1$,
remains in the interval $-2<b_1<2$, while the other, $b_2$, goes
 to negative values smaller than $-2$. For larger $E_j$
values, however, $b_2$ increases and for about $E_j \approx -0.224$,
 both indices
come in the stability zone, i.e. we have $-2<b_{1,2}<2$. The detailed
description of this complicated evolution is beyond the scope of the
present paper and does not add anything to the important information
that family x1v4 brings stable representatives in the system for $E_j
> -0.224$.  The x1v4 family remains stable up to $E_j \approx -0.149$ where it
becomes simple unstable. Its stability indices fold and the family continues
existing towards smaller energies.
The morphological evolution of x1v4 can be seen in
Fig.~\ref{orb-x1v4}. In Fig.~\ref{orb-x1v4}a we give the three
projections of an unstable orbit close to the bifurcating point from
which the family emanates, while in Fig.~\ref{orb-x1v4}b and
Fig.~\ref{orb-x1v4}c we give two stable orbits, for energies $E_j > -0.224$.
The last one is for $E_j =-0.206$ and we see that already the orbit reaches
$|z|$ values close to 2~kpc away from the equatorial plane. For each orbit of
this family there is also a symmetric one with respect to the equatorial
plane. If only one of the two is populated, this would give rise to an
asymmetric warp-like shape. Populating them both restitutes of course
symmetry.  The stable orbits of this family enhance the bar, but they deviate
substantially from the equatorial plane.

\subsection{Families at the  3:1 radial resonance}
\begin{figure}
\rotate[r]{    
\epsfxsize=6.0cm \epsfbox{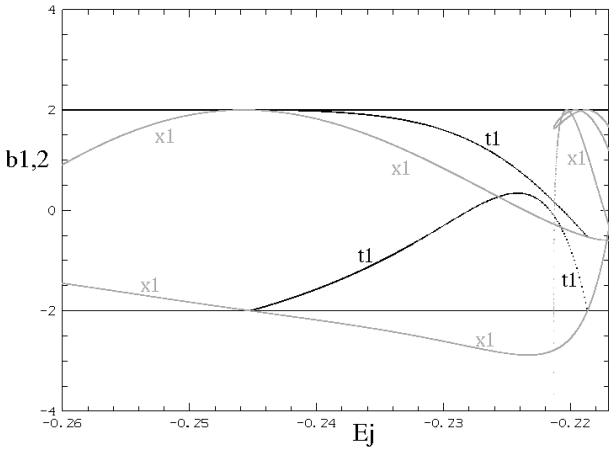}
}
\caption[]{Stability diagram of the t1 family, the first radial
  bifurcation of x1 at the 3:1 resonance. The stability indices of x1
  are given as well, drawn with light grey lines. }
\label{t1S}
\end{figure}

\begin{figure}
\rotate[r]{    
\epsfxsize=2.7cm \epsfbox{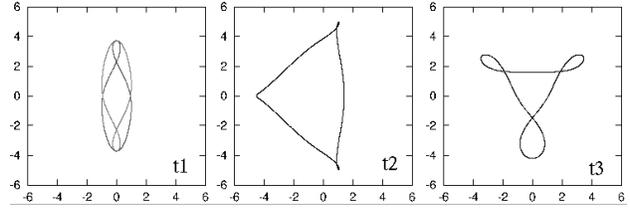}
}
\caption[]{Stable orbits of the three 2-dimensional families at the  3:1
  area. Note that the loops of t3 are asymmetric. In the left panel we
  plot a t1 orbit together with its symmetric with respect to the bar
  minor axis.} 
\label{o3:1-2D}
\end{figure}

\begin{figure}
\rotate[r]{ 
\epsfxsize=6.7cm \epsfbox{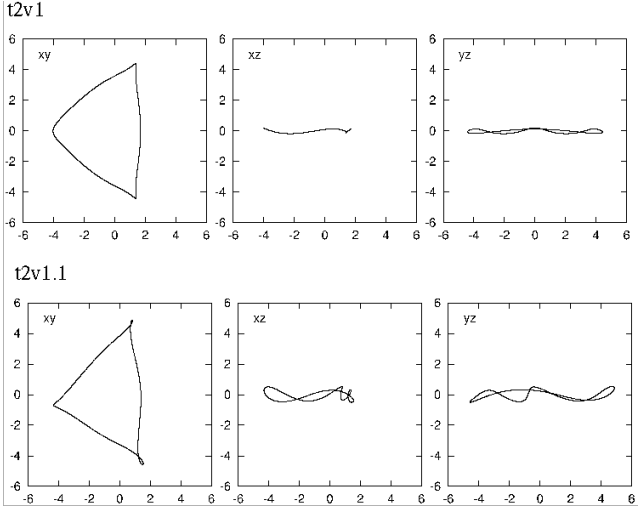}
}
\caption[]{Stable orbits of two 3-dimensional families at the 3:1
        area. Their names are given on the top left of each sets of
        panels.}
\label{o3:1-3D}
\end{figure}

As we already saw, one of the two stability indices bifurcated the 3D
families x1v1, x1v2, x1v3 and x1v4, by its intersections with the
$b=-2$ stability axis. The intersections of the second stability index
with this stability axis introduce in the system planar 2D orbits. The
first family is bifurcated after a S\ar U transition at $E_j \approx
-0.244$, i.e. in the 3:1 resonance region. We call it t1 and it is
stable (Fig.~\ref{t1S}).
It bridges exactly the instability zone of the x1 in the S\ar U\ar S
transition, i.e. its stability indices together with those of the x1,
form a bubble \cite{gco86}. t1 exists for approximately  $-0.244 < E_j
<-0.218$ and at $E_j \approx -0.218$ can be considered as an inverse
bifurcation\footnote{Inverse bifurcation is a non-linear phenomenon
encountered in Hamiltonian systems, according to which the bifurcated
family, instead of evolving towards the same direction as the parent
family, changes direction. It thus extends for the same energies as
the parent family before the transition and has the kind of stability
of the parent family after the transition. \cite{gco85}.} of x1.  At
$E_j \approx -0.214$, just beyond the `bow' area, the same stability
index has another intersection with the $b=-2$ axis and x1 bifurcates
another 2D family, t2. Several 2D and 3D 3:1 type families, related
to each other and with x1, are introduced in the interval
$-0.214<E_j< -0.20$.
Let us briefly mention that, besides t1 (in Fig.~\ref{t1S})
and t2, we found a third 2D 3:1 family, t3, which is stable for
$-0.2065 <E_j< -0.2005$, although it is introduced in the system as simple
unstable for $E_j \approx -0.205$. The
morphology of the three 2D families t1, t2 and t3 is given in
Fig.~\ref{o3:1-2D}, and their stable energy intervals in 
Table~\ref{tab:x1r}.
For the lower energies, the t1 orbits have only one loop, which is located
on the $y$ axis, as the example shown in the left panel of 
Fig.~\ref{o3:1-2D}. For higher energies they develop two more loops,
symmetric with respect to the $y$ axis, and roughly equal in size to the first
one. Since the orbits of family t1 are symmetric with respect to the
$y$ axis, for every orbit we should have also its symmetric with respect
to the $x$ axis. Combining the two, as in the left panel of
Fig.~\ref{o3:1-2D}, we obtain a shape that is elongated along the bar
major axis and resembles the morphology of the x1 orbits with loops,
at least for the energies where the orbits have only one loop. The
extent of such orbits along the $y$ axis reaches up to 4~kpc, i.e. two
thirds of the way to corotation.

t2 brings in the system three-dimensional families of periodic orbits
with stable representatives. It bifurcates the family t2v1
at $E_j \approx -0.209$, which in
turn bifurcates t2v1.1 at $E_j \approx -0.205$. 
The t2v1 family provides stable orbits to the system for $-0.209 < E_j <
-0.207$ and the t2v1.1 family for $-0.205 < E_j < -0.203$. Triangular-like
t2-type orbits have characteristic peaks at the sides of the bar, like the
peak of the orbits at $x \approx -4$ in the $(x,y)$
projections of Fig.~\ref{o3:1-3D}. They are
near but not always on the minor axis of the bar and their presence can lead to
local enhancements of the density at the area between the bar and the
L$_{4,5}$ points.
For any energy in the interval
$-0.214<E_j<-0.20$ there are almost always stable 3:1-type orbits of
one or the other family. Together with t1, they affect the dynamics of
the bar in this region. We note that
the 3:1 orbits bifurcated from x1 are very common in all barred
potentials and have both in 2D and 3D dynamically only local importance.
Orbits of type t2v1 and t2v1.1 have been found even in the early
$N$-body simulations of 3D bars (Figure 5 in Miller \& Smith
1979). The loops of t3 on either side of the major axis of the bar are
not of equal size. As can be seen by careful inspection of the t3
orbit in Fig.~\ref{o3:1-2D}, the loop at the right side of the major
axis is slightly bigger than the one to the left. Thus,
morphologically, t3 is a kind of asymmetric t1, since for larger
energies t1 develops loops which are symmetric with respect to the
major axis, besides the one along the major axis.

\subsection{The last part of the x1-tree}
\begin{figure}
\rotate[r]{
\epsfxsize=3.0cm \epsfbox{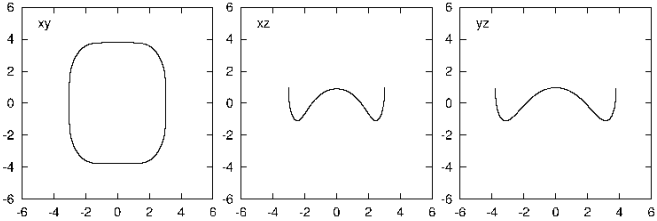}
}
\caption[]{A typical stable x1v5 orbit. }
\label{ox1v5}
\end{figure}

\begin{figure}
\rotate[r]{ 
\epsfxsize=5.4cm \epsfbox{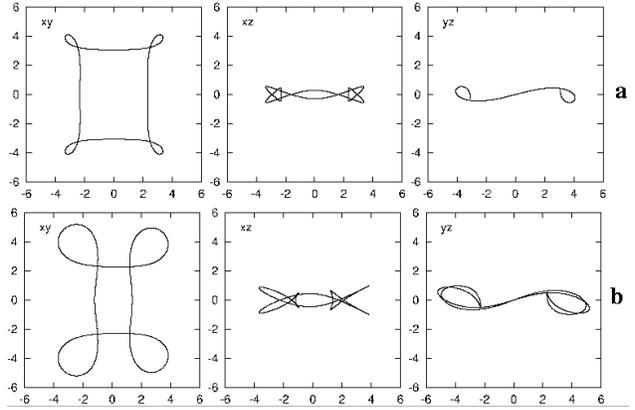}
}
\caption[]{(a) A x1v7 orbit and (b) a x1v7.1 one. Both are stable.
 }
\label{ox1v7}
\end{figure}

There are two more 3D bifurcations of x1 close to the local maximum of
the characteristic at $E_j \approx -0.205$. It is x1v5 (bifurcated at
$E_j \approx -0.213$ and being stable until $E_j \approx -0.172$) and x1v7
(bifurcated at $E_j \approx$ -0.205, just beyond the peak of the
characteristic). The family x1v7 and its bifurcation x1v7.1 provide
stable orbits for $-0.205<E_j<-0.18$ and
$-0.175<E_j<-0.17$. Nevertheless, the part of this family that
contributes to the density of the bar is limited by the fast increase
of $|z|$ with the energy. Fig.~\ref{ox1v5} and \ref{ox1v7} 
show the morphology of these families.

In the same region we encounter two more 3D bifurcations of x1, namely
the families x1v6, introduced at $E_j \approx -0.211$
(Fig.~\ref{x1-fiogkos}), and x1v8 introduced at $E_j \approx -0.1925$
(Fig.~\ref{x1-III}). Both are born after an U\ar S transition of x1
and remain always unstable. We note that the representatives of x1v5, x1v6,
x1v7 and x1v8 families are morphologically similar to those of 
the B$z_2$, B$\dot{z}_2$,
B$\dot{z}_3$ and B$z_3$ families of Pfenniger (1984) respectively.

\begin{figure}
\rotate[r]{     
\epsfxsize=5.0cm \epsfbox{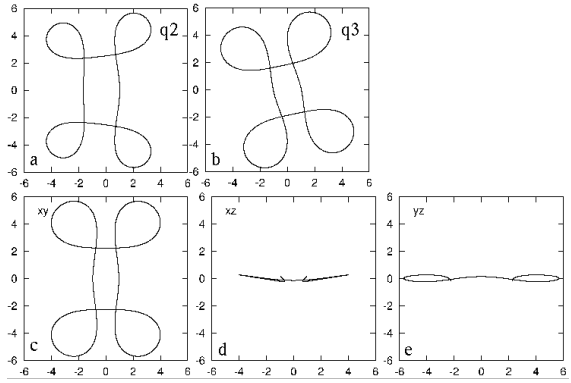}
}
\caption[]{Morphology of stable orbits of x1's bifurcations at largest
  energy values. Panels (a) and (b) show members of families q2 and q3
  respectively. Panels (c) to (e) show the three views of an orbit of
  the x1v9 family.}
\label{o-x1tel}
\end{figure}

As we have seen, x1 is mostly unstable in the decreasing branch beyond
the local maximum at the radial 4:1 gap and the morphology of the
orbits at this branch is in general rectangular-like with loops in the
corners. There are several families bifurcating from this branch and
their orbits have, as already noted for other families, a morphology
similar to that of x1 in the region. The 2D families q2 and q3 provide
stable asymmetric orbits, two examples of which are given in
Fig.~\ref{o-x1tel}. We also have one 3D bifurcating family, x1v9, a
member of which is shown in Fig.~\ref{o-x1tel}. This family also has
an asymmetric stable bifurcation for a short energy interval. No stable
members of these families can be found outside the interval
$-0.186<E_j<-0.1808$. To this we should add the small intervals of
stability provided by x1 itself (cf. Fig.~\ref{x1char} and
Fig.~\ref{x1-III}).

\begin{figure}
\rotate[r]{     
\epsfxsize=8.0cm \epsfbox{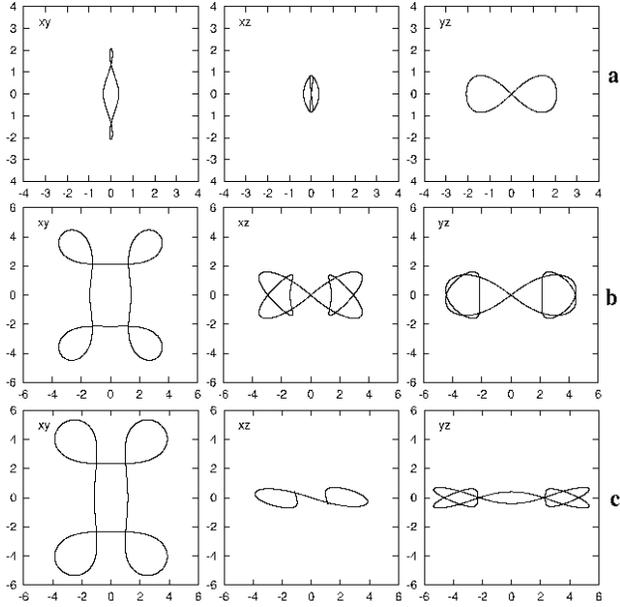}
}
\caption[]{Orbits of the unstable families x1v2 (a), x1v6 (b), and x1v8 (c)
at $E_j= -0.2612,\; -0.178$ and $-0.184$ respectively.}
\label{unstorb}
\end{figure}

Finally, for the sake of completeness we give in Fig.~\ref{unstorb}
the morphology of the three 3D families, members of the x1-tree, which
remain always unstable although they exist for large energy
intervals. As we have seen in the corresponding paragraphs they are
the families x1v2, x1v6 and x1v8.

\section{Further families}
\subsection{Orbits around L$_4$ and L$_5$}
\begin{figure}
\rotate[r]{     
\epsfxsize=8.0cm \epsfbox{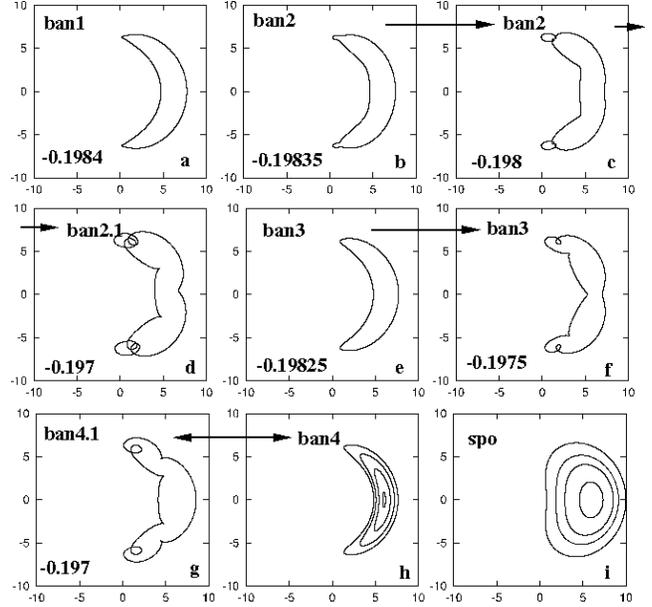}
}
\caption[]{Stable 2D banana-like orbits. $E_j$ is given in the lower
  left corner of panels (a) to (g). Panels (h) and (i) include orbits
  of many $E_j$ values. Arrows indicate morphological evolution of the
  same or related families.}
\label{ban2d}
\end{figure}

Another important `forest' of families is the group of the banana like
orbits. Here we find the usual planar long and short period
banana-like orbits \cite{cg89}. The long period orbits are coming in
the system in a large variety of families all of which have stable
parts for $-0.1984 < E_j < -0.1944$. 
The stability indices of these orbits exhibit a complicated behaviour  having
several tangencies and intersections with the $b=2$ and $b=-2$ axes. This
brings many families in the system by bifurcation. 
The family found for lowest $E_j$ values is
ban1 (Fig.~\ref{ban2d}a) which is born at $E_j \approx -0.1984$, followed by
ban2 (Fig.~\ref{ban2d}b,c) that appears at a slightly greater energy value -
which in turn bifurcates ban2.1 (Fig.~\ref{ban2d}d) at $E_j \approx -0.1972$ -
and ban3 (Fig.~\ref{ban2d}e,f) introduced in the system at $E_j \approx
-0.1983$. The most important of the planar orbits with stable parts, is ban4
(Fig.~\ref{ban2d}h), because it is stable over the largest energy interval
($-0.1982 < E_j < -0.1955$). It exists for $E_j > -0.1982$ and it is not
bifurcated at this point from any of the families 
existing for lower energies (ban1, ban2, ban2.1, ban3, ban3.1). From ban4
bifurcates the 2D family ban4.1 (Fig.~\ref{ban2d}g). The stability indices of
ban4 have a complicated behaviour which is typical of a collision of
bifurcations \cite{gco86}\footnote{Collisions of bifurcations happen when both
$b_1$ and $b_2$ are exactly equal to $-2$ or $2$ for a particular set
of the control parameters. In order to observe a collision we need to
vary continuously a control parameter of our model (i.e. to consider
successive individual models), and for all these cases to follow the
evolution of the stability indices as a function of $E_j$. This
practically means that we vary {\em two} parameters. If it happens
that $b_1 = b_2 = -2$ (or 2) for a critical set of the control
parameters, then we will observe a change in the interconnections
between parent and bifurcating families, before and after the
collision. This may also change the general behaviour of the dynamical
system.}. Approaching $E_j =-0.1955037765$, the ban4 orbits shrink to
L$_4$ (or L$_5$), and beyond this point the short period orbits (spo)
grow in size and take their bean-like shape (Fig.~\ref{ban2d}h and i
respectively).

\begin{figure}
\rotate[r]{  
\epsfxsize=8.0cm \epsfbox{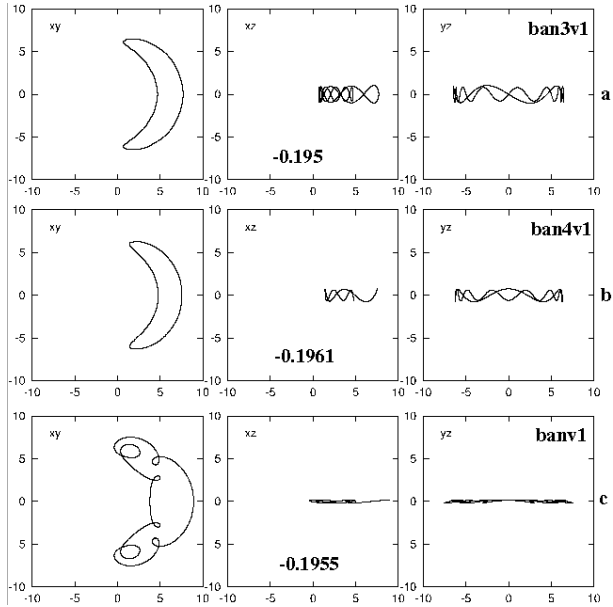}
}
\caption[]{Stable 3D banana-like orbits. All three extend to large
  $E_j$ values but the two most important (ban3v1, ban4v1) are at
  these large $E_j$ values complex unstable. The numbers at the bottom
  of the $(x,z)$ projections give the $E_j$ of each orbit.}
\label{ban3d}
\end{figure}

We have also found three 3D families of periodic orbits with stable
parts.  ban3v1 (Fig.~\ref{ban3d}a), a bifurcation of ban3 at $E_j \approx
-0.1982$, is initially marginally stable, having one of its two stability
indices almost equal to $-2$, but for $E_j > -0.1962$ the index become
clearly larger than $-2$. At $E_j \approx -0.1947$ the two indices join each
other and we have a S\ar \D transition.   ban4v1 (Fig.~\ref{ban3d}b), a
bifurcation of ban4 at $E_j \approx -0.1976$, is almost everywhere marginally
stable in the interval $-0.1976 < E_j < -0.1944$.  For $E_j > -0.1944$ it is
always complex unstable.  ban3v1 and ban4v1 extend to very large $E_j$ values,
but as complex unstable. Since we have S\ar \D transitions there are no
bifurcating families and this is the mechanism that terminates the trapping of
material around banana-like orbits in our 3D bars. Finally banv1 (Fig.~\ref{ban3d}c)
is introduced in the system at $E_j \approx -0.1957$ as stable and remains
stable up to $E_j \approx -0.1954$. This family is not obviously related to
any other banana-like orbit. Since it is a 3D family we name it banv1.
\subsection{Orbits around L$_1$ and L$_2$}
\begin{figure}
\rotate[r]{ 
\epsfxsize=4.0cm \epsfbox{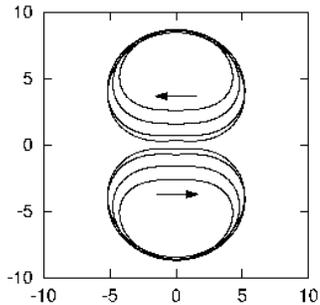}
}
\caption[]{Stable orbits of the $\ell_1$ family and their symmetrics with
respect to the $x$ axis. The innermost orbit, just next to the arrow,
corresponds to $E_j \approx$ -0.168, just after the U\ar S transition.
}
\label{ellorb}
\end{figure}

The L$_1$, L$_2$ Lagrangian points are known to be always unstable
\cite{bt87}. Around them we find a family of planar periodic orbits we
call $\ell_1$. It appears for $E_j$ values larger than the one
corresponding to L$_1$, the morphology of its orbits resembles that of
the spo orbits rotated by $\pi/2$, and their periods are of the order
of the epicyclic period. Close to the L$_1$ energy and for $E_j<
-0.168$ these orbits are unstable. For $E_j >-0.168$, however,
$\ell_1$ has both stability indices between $-2$ and 2 and the family
becomes stable. Orbits of this family can be found only by starting with
initial conditions on the major axis of the bar. For this reason they
had not been previously found, since in previous studies searches for
periodic orbits started only with initial conditions on the $y=0$
axis. In Fig.~\ref{ellorb}, we plot a few stable orbits of $\ell_1$
and their symmetrics with respect to the $x$ axis for $E_j >-0.168$.
These stable orbits do not support the bar since they are elongated
parallel to the minor axis. Nevertheless, they are of physical
interest since they support motion parallel to the minor axis,
contribute to the exchange of material between regions inside and
outside corotation and are able to influence the dynamics in the
region between bar and spirals in barred spiral galaxies. The
streaming at the apocentra of the $\ell_1$ orbits could support
arc-like features beyond the end of the bar.

For larger energies the $\ell_1$ orbits can be observed shifted
towards the $x$ axis (minor axis of the bar), at about $E_j \approx
-0.12$ they cross the $x$ axis and after a short unstable zone they fall
on the retrograde family x4 as stable.

\subsection{Orbits outside corotation}
\begin{figure}
\rotate[r]{  
\epsfxsize=2.5cm \epsfbox{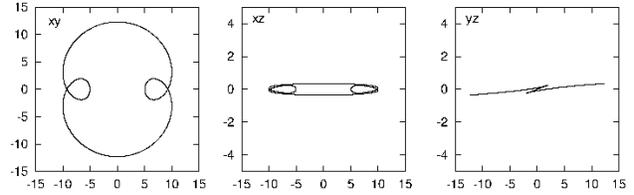}
}
\caption[]{A 3D stable orbit beyond corotation.
 }
\label{3Doc}
\end{figure}

Beyond corotation we find the usual planar families \cite{cg89}. Most of their
members display 
loops. We also find several 3D families with stable parts. As an
example we give the family depicted in Fig~\ref{3Doc}, which is a
bifurcation of the planar family called x1(1) by Contopoulos \&
Grosb{\o}l (1989). The vertical extent of the 3D orbits we found
beyond corotation is in general small.

\begin{figure}
\rotate[r]{    
\epsfxsize=5.5cm \epsfbox{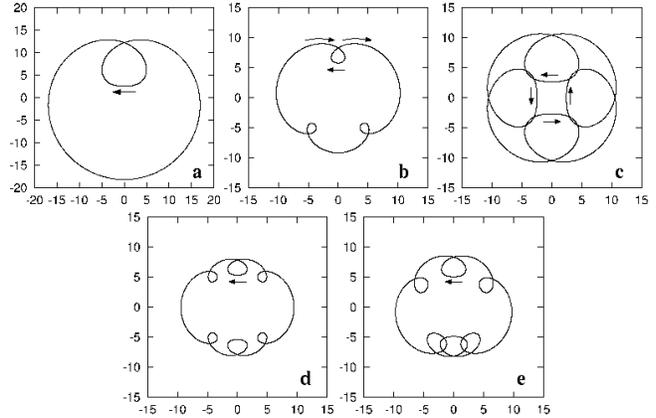}
}
\caption[]{Examples of stable 2D orbits beyond corotation. They
support motion parallel to the minor axis of the bar at the corotation
region and favour exchange of material between regions far from the
center and regions near to it.}
\label{olr-pan}
\end{figure}

Let us also mention some 2D families, orbits of which are given in
Fig.~\ref{olr-pan}. They have been calculated starting with initial
conditions on the major axis of the bar as in the case of the $\ell_1$
family and have thus not been described in previous papers.  All of
them have large stable parts. These orbits have two interesting
properties.  First, some of them could support motion close to the end
of the bar parallel to its minor axis, at radii shorter than the
corotation radius. Second, they could efficiently transport material
from the outer parts of the disc, e.g.  from a distance close to
20~kpc from the center, to the central regions of the bar
(e.g. Fig.~\ref{olr-pan}a). This is particularly true for orbits as
those shown in Fig.~\ref{olr-pan}a and c.

\section{Conclusions}
In this paper we have made an extensive study of both the 2D and 3D periodic
orbits in a fiducial model representative of a barred galaxy. We report on the
stability and morphology of the main families.  Our main conclusions are:

\begin{enumerate}
 
\item So far the x1 orbits were considered the backbone of bars. This,
  however, can only be the case for 2D bars, since the x1 can only
  populate the $z$=0 plane. For 3D bars the backbone is the x1, together
  with the tree of its 3D bifurcating families. Trapping around these
  families will determine the thickness and the vertical shape of
  galaxies in and around the bar region.  Major building blocks for
  the 3D bars can be supplied also by families initially introduced as
  unstable. Thus the family x1v4, introduced in the system after a U\ar S
  transition, is a basic family, giving stable representatives
  for large energy intervals in the system.

\item The $(x,y)$ projections of the 3D families of the x1-tree retain
  in general a morphological similarity with their parent family
  at the same energy. This has important implications for the
  morphology of a galaxy since it introduces building blocks which
  have similar morphology as the x1 orbits, but have a considerable
  vertical extensions. Especially at the regions close to the
  bifurcating points the $(x,y)$ morphology of a x1v$n$ family is not
  only geometrically similar, but actually very close to the 
  morphology of the corresponding x1 orbit.
  
\item The way the 3D families of the x1-tree are introduced in the system at
  an instability strip determines the importance of the bifurcations in $z$ or
  $\dot{z}$. Particularly in the present model, all 3D families of the x1-tree
  at the increasing part of the characteristic which are bifurcated in $z$ are
  introduced in the system as stable. On the other hand the stable family
  associated with the 5:1 vertical resonance (x1v7), bifurcated at the
  decreasing part of the x1 characteristic, beyond its local maximum, is
  bifurcated in $\dot{z}$. Whether the stable family of the last S\ar U\ar S
  transition is the bifurcation in $z$ or $\dot{z}$ determines in a large
  degree the model's morphology at its outer parts.

\item The radial 3:1 resonance region provides in the system several
  2D and 3D stable families. Their role, however, is locally confined,
  as in 2D models.

\item 3D orbits elongated along the minor axis of the bar can be given by
  bifurcations of the planar x2 family. 


\item We have found several families of 3D banana-like orbits around
  L$_{4,5}$. Their extent is always restricted by a S\ar \D
  transition.

\item Stable families found beyond corotation circulate material
  between the outer parts of the system and regions as far inwards as
  1~kpc.  This contributes to the mixing of the elements in a disc
  galaxy.

\end{enumerate}

The families of periodic orbits we described up to now are indeed the
basic families of a 3D Ferrers bar. As we explore the parameter
space, however, their properties change, while new important families
may appear and play a crucial role. A notable example is z3.1s, a
family related to the z-axis orbits along the rotational axis, which
will be described in paper II. However, these are rather particular
cases and are not encountered in every model.

\section*{Acknowledgments}
We acknowledge fruitful discussions and very useful comments by
Prof.~G.~Contopoulos. 
We thank the referee for useful suggestions that allowed to improve the
presentation of our work.
This work has
been supported by E$\Pi$ET II and K$\Pi\Sigma$ 1994-1999;
and by the Research Committee of the Academy of Athens. Ch.~Skokos and
P.A.~Patsis thank the Laboratoire d'Astrophysique de Marseille for an
invitation during which essential parts of this work have been completed.

\bsp

\label{lastpage}


\begin{thebibliography}{}
\bibitem[Athanassoula ] {ath84} Athanassoula  E., 1984 Phys. Rep. 114, 319

\bibitem [Athanassoula 1992a] {ath92a} Athanassoula  E., 1992a, MNRAS 259, 328

\bibitem [Athanassoula 1992b] {ath92b} Athanassoula  E., 1992b, MNRAS 259, 354

\bibitem [Athanassoula 1996] {ath96} Athanassoula  E., 1996, in `Spiral
                      Galaxies in the near-IR', D. Minniti \& H.-W. Rix (eds.),
                      p. 147, Springer

\bibitem[Athanassoula et al. 1983] {ath83} Athanassoula
  E., Bienayme O., Martinet L., Pfenniger D., 1983, A\&A 127, 349

\bibitem [Binney \& Tremaine 1987] {bt87} Binney J., Tremaine S.,
1987, `Galactic Dynamics', Princeton University Press, Princeton, N.J.

\bibitem [Broucke 1969] {br} Broucke R., 1969, NASA Techn. Rep. 32, 1360

\bibitem[Combes et al. 1990] {cetal90} Combes F., Debbasch
  F., Friedli D., Pfenniger D., 1990, A\&A 233, 95  

\bibitem [Contopoulos 1980] {gco80} Contopoulos G., 1980, A\&A 81, 198

\bibitem [Contopoulos 1985] {gco85} Contopoulos G., 1985, in `Chaos in
  Astrophysics', J.R. Buchler et al (eds.), p. 259, Reidel

\bibitem [Contopoulos 1986] {gco86} Contopoulos G., 1986, Celest. Mech. 38, 1

\bibitem [Contopoulos \& Barbanis 1985] {cb85} Contopoulos G.,
          Barbanis B., 1985, A\&A 153, 44

\bibitem [Contopoulos \& Grosb{\o}l 1988] {cg88} Contopoulos G., Grosb{\o}l
  P.,   1988, A\&A 197,83

\bibitem [Contopoulos \& Grosb{\o}l 1989] {cg89} Contopoulos G., Grosb{\o}l P.,
  1989, A\&AR, 1,261

\bibitem [Contopoulos \& Magnenat 1985] {cm} Contopoulos G., Magnenat P.,
  1985,   Celest. Mech. 37, 387 

\bibitem [Contopoulos \& Mertzanides 1977]{cme}  Contopoulos G., Mertzanides
  C., 1977, A\&A 61, 477

\bibitem [Hadjidemetriou 1975] {ha} Hadjidemetriou J., 1975,
  Celest. Mech. 12,  255  

\bibitem [Hasan, Pfenniger \&  Norman 1993] {hpfn93} Hasan H., Pfenniger D.,
  Norman C., 1993, ApJ   409, 91

\bibitem [Heggie 1985] {dh85} Heggie D.C., 1985, Celest. Mech. 35, 357

\bibitem [Heisler et al. 1982] {hms} Heisler J., Merritt D., Schwarzschild
 M., 1982,  ApJ  258, 490

\bibitem [Kormendy 1982] {Kor82} Kormendy J., 1982, in `Morphology and 
 Dynamics of Galaxies', L. Martinet and M. Mayor eds., 12th Advanced
 Course, Saas-Fee, p. 113, Geneva Obs., Sauverny

\bibitem [Martinet \& Pfenniger 1987] {mpf87} Martinet L., Pfenniger D., 1987,
  A\&A 173, 81
\bibitem [Martinet \& de Zeeuw 1988] {mdz88} Martinet L., de Zeeuw T., 1988,
  A\&A 206, 269

\bibitem[Miller \& Smith 1979] {ms79} Miller R.H., Smith B.F., 1979, ApJ 1979,
  227, 785

\bibitem [Miyamoto \& Nagai 1975] {mina 75} Miyamoto M., Nagai R., 1975, PASJ
  27, 533 

\bibitem [Olle \& Pfenniger 1998] {opf98} Olle M.,  Pfenniger D., 1998 A\&A
  334, 829

\bibitem [Patsis et al. 1997] {paq} Patsis P.A., Athanassoula  E., Quillen
  A.C., 1997, ApJ 483, 731

\bibitem [Patsis \& Grosb{\o}l 1996] {pg96} Patsis P.A., Grosb{\o}l P., 1996,
A\&A 315, 371  

\bibitem [Patsis et al. 2002] {p3} Patsis P.A., Skokos Ch., Athanassoula
   E., 2002 ({\em paper III - in preparation})

\bibitem [Patsis et al. 2002] {p4} Patsis P.A., Skokos Ch., Athanassoula
   E., 2002 ({\em paper IV - in preparation})

\bibitem [Pfenniger 1984] {pf84} Pfenniger D., 1984, A\&A 134, 373

\bibitem [Pfenniger 1985a] {pf85a} Pfenniger D., 1985a, A\&A 150, 97

\bibitem [Pfenniger 1985b] {pf85b} Pfenniger D., 1985b, A\&A 150, 112

\bibitem [Pfenniger 1987] {pf87} Pfenniger D., 1987, A\&A 180, 79

\bibitem [Pfenniger 1990] {pf90} Pfenniger D., 1990, A\&A 230, 55

\bibitem [Pfenniger 1996 ] {pf96} Pfenniger D., 1996, in `Barred Galaxies',
  ed. R. Buta, D. A. Crocker and B. G. Elmegreen, ASP  Conf. Ser. 91, p.273

\bibitem [Poincar\'{e} 1899] {poin} Poincar\'{e} H, 1899, `Les
  Methodes Nouvelles 
  de la Mechanique Celeste', Vol III, Gauthier-Villars, Paris     

\bibitem [Polymilis et al., 1997] {poly} Polymilis C., Servizi G.,
Skokos Ch., 1997, Celest. Mech. Dyn. Astron. 66, 365

\bibitem [Sellwood \& Wilkinson 1993] {sw}  Sellwood J., Wilkinson A., 1993,
  Rep. Prog. Phys. 56, 173
 
\bibitem [Schwarzschild, 1979] {schw} Schwarzschild M., 1979, ApJ 232,236

\bibitem [Skokos 2001]{sk01} Skokos Ch., 2001, Physica D 159, 155 

\bibitem [Skokos et al. 2002]{skea2}  Skokos Ch., Patsis P.A., Athanassoula
   E., 2002 MNRAS - this issue 
\end{thebibliography}
\end{document}